\documentclass[12pt,reqno]{amsart}
            
\usepackage{geometry}                
\usepackage{graphicx,color,bm,float}
\usepackage{amsmath,amssymb,amsthm,amsfonts} 
\usepackage{bbm}
\usepackage{times}
\usepackage{cite}
\usepackage{enumerate}
\usepackage{bbm}
\usepackage{url}
 \newcommand{\beq}{\begin{equation}}
  \newcommand{\eeq}{\end{equation}}
  \newcommand{\beql}[1]{\begin{equation}\label{eq:#1}}

  \newcommand{\beqa}{\begin{eqnarray}}
  \newcommand{\eeqa}{\end{eqnarray}}
  \newcommand{\beqas}{\begin{eqnarray*}}
  \newcommand{\eeqas}{\end{eqnarray*}}
    \newcommand{\bal}{\begin{align}}
  \newcommand{\eal}{\end{align}}
  \newcommand{\bals}{\begin{align*}}
  \newcommand{\eals}{\end{align*}}
  \newtheorem{Theorem}{Theorem}[section]

 \newenvironment{Proof}{\begin{trivlist}
    \item[\hskip \labelsep {\em \indent Proof.}]}{\qed\end{trivlist}}

  \newcommand{\C}{\mathbf{C}}
  \newcommand{\al}{\alpha}
 \newcommand{\av}[1]{\left\langle#1\right\rangle}
  \newcommand{\be}{\beta}
  \newcommand{\da}{\ast}
  \newcommand{\de}{\delta}
  \newcommand{\ga}{\gamma}
  \newcommand{\nn}{\nonumber}
  \newcommand{\om}{\omega}
  \newcommand{\rh}{\rho}

  \newcommand{\Om}{\Omega} 
  \newcommand{\Eq}[1]{Eq.~(\ref{eq:#1})}
  \newcommand{\Tr}{{\rm Tr}}
 \newcommand{\ba}{\mathbf{a}}
 \newcommand{\bb}{\mathbf{b}}
  \newcommand{\bx}{\mathbf{x}}

  \newcommand{\eq}[1]{(\ref{eq:#1})}
\newcommand{\bra}[1]{\langle#1|}
\newcommand{\ket}[1]{|#1\rangle}
\newcommand{\ketbra}[1]{\ket{#1}\bra{#1}}

\newcommand{\bmat}{\left[\begin{array}{rr}}
\newcommand{\emat}{\end{array}\right]}
\newcommand{\bvec}{\left[\begin{array}{r}}
\newcommand{\evec}{\end{array}\right]}
\newcommand{\btmat}{\left[\begin{array}{rrr}}
\newcommand{\etmat}{\end{array}\right]}
\newcommand{\cA}{{\mathcal A}}
  \newcommand{\cB}{{\mathcal B}}
  \newcommand{\cH}{{\mathcal H}}
  \newcommand{\cI}{{\mathcal I}}
  \newcommand{\cK}{{\mathcal K}}
  \newcommand{\cL}{{\mathcal L}}
  \newcommand{\cT}{{\mathcal T}}
\newcommand{\benum}{\begin{enumerate}[{\rm (i)}]\itemsep=0in}
\newcommand{\eenum}{\end{enumerate}}
\newcommand{\kps}{\ket{\psi}}
\newcommand{\kph}{\ket{\varphi}}
\newcommand{\kxi}{\ket{\xi}}

\newcommand{\Perp}{^{\perp}}
\newcommand{\p}{^{\perp}}
\newcommand{\bProof}{\begin{proof}}
\newcommand{\eProof}{\end{proof}}
\newcommand{\bTheorem}{\begin{Theorem}}
\newcommand{\eTheorem}{\end{Theorem}}
\newcommand{\deq}[1]{\begin{align}#1\end{align}}
\newcommand{\deqs}[1]{\begin{align*}#1\end{align*}}
\newcommand{\deqed}[1]{\begin{equation}\begin{aligned}#1\end{aligned}\end{equation}}
\newcommand{\q}{\bar{p}}
\newcommand{\id}{{\rm id}}
\renewcommand{\star}{\ast}

\title[Modeling QOE, RRE, and QQE with quantum instruments]
{Modeling combination of question order effect, response replicability effect, and QQ-equality with quantum instruments}
\author[M. Ozawa and A. Khrennikov]{Masanao Ozawa\\
College of Engineering, Chubu University, 1200 Matsumoto-cho, Kasugai 487-8501, Japan\\
Graduate School of Informatics, Nagoya University, Chikusa-ku, Nagoya 464-8601, Japan\\
Andrei Khrennikov\\ 
Linnaeus University, International Center for Mathematical Modeling\\  in Physics and Cognitive Sciences
 V\"axj\"o, SE-351 95, Sweden}
\date{}                     
\begin{document}
\begin{abstract} We continue to analyze basic constraints on the human decision making from the viewpoint 
of quantum measurement theory (QMT). As it has been found, the conventional QMT  based on the projection postulate 
cannot account for the combination of the question order effect (QOE) and the 
response replicability effect (RRE). This was alarming finding for quantum-like modeling of decision making. Recently, it was shown 
that this difficulty can be resolved by using of the general QMT based on quantum instruments. 
In the present paper we analyse the problem of the combination of QOE, RRE, and the well-known QQ-equality (QQE). 
This equality was derived by Busemeyer and Wang and it was shown (in a joint paper with Solloway and Shiffrin) that statistical data from many social opinion polls
 satisfy it.  Here we construct quantum instruments  satisfying QOE, RRE and QQE.
The general features of our approach are formalized with postulates that generalize (the Wang-Busemeyer) postulates for quantum-like modeling of decision making. Moreover, we show that our model closely reproduces the statistics of the well-known
Clinton-Gore Poll  data  with a prior belief state independent
of the question order.  This model successfully corrects for the order
effect in the data to determine the ``genuine'' distribution of the opinions in
the Poll. The paper also provides an accessible
introduction to the theory of quantum instruments - the most general mathematical framework for quantum measurements. 
\end{abstract}

\maketitle

keywords: decision making, quantum instruments, quantum-like models,  order effect,  QQ-equality, response replicability effect, social science   

\section{Introduction} 
There has been a growing interest in cognitive psychology, behavioural economics, political sciences, and molecular biology in quantum formalism applications (e.g. books \cite{QL0zz}-\cite{QL5zz}). The majority of applications are based on quantum measurement theory (QMT). However, as shown in \cite{PLOS}, the conventional QMT based on the projection postulate (von Neumann-L\"uders \cite{VN,Luders})  representing the mental state update has difficulties in confronting with combination of some  psychological effects. (Classical Bayesian update based on Kolmogorov probability \cite{K} has difficulties even with individual effects, see \cite{POL}-\cite{BAX}.)  So, QMT may describe each of the effects individually, but not jointly. In \cite{PLOS}, it was shown that combination of {\it the question order effect} (QOE)\footnote{QOE: dependence of the (sequential) joint probability distribution of answers on the questions' order: $p_{AB}\not=p_{BA};$ see \cite{Moo02} for experimental statistical data (the Clinton-Gore social opinion poll) and  \cite{WB13,WSSB14} for its modeling with conventional QMT.} 
and {\it the response replicability effect }(RRE)\footnote{RRE \cite{PLOS}: Suppose that after answering the $A$-question with 
``yes'', Alice is  asked another question $B, $ and gives an answer to it. 
And then she is asked $A$ again. In the social opinion polls and other decision making experiments, Alice would definitely repeat her original answer to $A,$ ``yes''. This is $A-B-A$ response replicability. (In the absence of $B$-question, we get $A-A$ replicability). Combination of $A-B-A$ and $B-A-B$ replicability forms RRE.} cannot be described by the conventional QMT.  This was an alarming finding for quantum-like modeling of decision making.\footnote{Contrary to QOE, RRE has not been discussed as much in psychological literature. The role of this effect in testing the boundaries of applicability of the quantum formalism in psychology was  emphasized in paper \cite{PLOS}. In this paper, the Clinton-Gore opinion poll (see Moore \cite{Moo02})  was considered as the main example. Although in such social opinion polls one can expect the combination of QOE+RRE, it is very important to perform detailed 
experimental studies. The first attempt to proceed with such experimentation was made in paper of Busemeyer and Wang \cite{BWRRE}. 
This paper generated critical comments  \cite{KRRE1}-\cite{DRRE} and, to clarify the situation,  further experiments have to be performed.}

However, since the 1970s QMT has been further developed towards more flexible approach to treating all the physically realizable
quantum measurements by abandoning the projection postulate of the conventional QTM,  
used during the first years of quantum physics. Nowadays,  in quantum physics, especially in quantum information theory, one uses QMT based on the theory of quantum instruments \cite{DL70}-\cite{OO}.\footnote{The generalization of the notion of quantum observables in the form of positive operator valued measures (POVMs) (which are widely used in quantum information theory and begin to be used in quantum-like modeling \cite{QL1,POVM}) appears naturally in the framework of theory of quantum instruments (see Appendix \ref{apovm}). However, we shall not use POVMs in the present paper.} Thus in quantum information theory, it would not be surprising that in some situations  conventional QMT (based on the projection postulate) cannot be  applied. 

Therefore, the right reply to the challenge presented in  \cite{PLOS} should be based on the theory of quantum instruments. The first 
attempt to proceed in this direction was made in the paper of Basieva and Khrennikov \cite{FOUND}. Still, they used a restricted class of quantum instruments. (This class is standard for quantum information theory.) Their paper confirmed the impossibility statement formulated originally 
in \cite{PLOS}. Only very recently it was shown \cite{ENTROPY} that by using quantum instruments, it is possible to describe the combination of QOE and RRE.  Thus, the quantum-like modeling program was released from the objection raised in paper \cite{PLOS}. 

Immediately after publication of article  \cite{ENTROPY}, Dzhafarov (private communication)  raised the question whether the model presented in \cite{ENTROPY} matches the QQ-equality (QQE). The latter is one of the most important results of the quantum-like approach to decision making. This is a special constraint  on probabilities that was derived by Busemeyer and Wang  \cite{WB13} in the quantum probabilistic framework. So, from the classical probabilistic  viewpoint there is no reasons for QQE to hold, but usage of quantum formalisms may require it. In paper  \cite{WSSB14},   it was shown that statistical data from a family of social opinion polls satisfies QQE. We stress that this equality was derived on the basis of conventional QMT \cite{WB13} and that  it can be violated in general QMT \cite{DL70}-\cite{OO}. Thus, combination of QOE and RRE with QQE is a delicate problem: one has to go beyond conventional QMT in a special way.

The purpose of this paper is two-fold:

A). We present the solution of this problem (QOE+RRE+QQE) by constructing the corresponding quantum instruments. This step is important in order to justify the  use of the quantum-like models for decision making. The general features of our approach are formalized with postulates that generalize Wang-Busemeyer postulates for quantum-like modeling of decision making \cite{WB13}. 

We  emphasize that our model is a complete model for a certain set of data satisfying the QQ-equality, including the
Clinton-Gore poll  (see \cite{Moo02}), in the following sense. 
Since the Clinton-Gore poll data sets approximately satisfy QQE, there should be a uniform method to make the original data 
to satisfy QQE with small distortion.   Here the main point is that {\it such a method can be defined independent of the model.}  
We show that our model completely reproduces the modified data of  the Clinton-Gore poll data and this means our model 
reproduces the original data as accurately as the data fit QQE.
Moreover, we show that this model closely reproduces the statistics of the 
Clinton-Gore poll data with a prior belief state independent
of the question order.  The model successfully corrects for the order
effect in the data.
\footnote{It should be pointed out that we are not trying to compare the classical \cite{K} and  quantum  \cite{VN} probability models applied for decision making. We remark  that in quantum physics comparison of these models began in the 1920s (with the Wigner function) and is  still continuing, e.g., in the discussions on Bell's inequality \cite{B1,B2,B3}. How far can one proceed with classical probability in quantum physics? This is a very complex problem (see, e.g., \cite{CP7}-\cite{CP77}) and it is too early to make a final conclusion.   Nevertheless, the quantum formalism has been successfully applied to numerous theoretical and engineering problems. This formalism is powerful and successful, irrespective of  the (im)possibility to proceed with classical probability.
One of the advantages of quantum formalism is linearity of the state space. This reduces required calculations to simple linear algebra for matrices and vectors. In fact, the linear space structure of mental spaces has been widely used in cognitive science and psychology. Quantum-like modeling continues the tradition of  modeling cognitive processes with linear state spaces, with the advantage of enhanced mathematical formalism and methodology  elaborated in  quantum physics, one of the most successful branches of science.}

In this paper we do not discuss alternative theories. We only offer quantum models as viable candidates for the considered experiments,
without comparison with possible other state-of the art models conventionally used in psychology. For such discussion we refer the reader to works  \cite{B1,B2,B3}.

B). Irrespective of the (QOE+RRE+QQE)-problem, the paper presents  the most general formalism 
of QMT based on theory of quantum instruments.   We hope that this theory will attract attention of psychologists and stimulate further applications in psychology.

Finally, we remark that the projective type  (von Neumann-L\"uders \cite{VN,Luders}) instruments, i.e., describing the measurement's feedback onto the system's state by orthogonal projections (section \ref{VNL}), are widely used in quantum-like modeling in cognition and 
decision making \cite{BB12,WB13}, \cite{WSSB14}-\cite{BAX}, \cite{White1,White2} (cf. with more complex models \cite{POVM,BAG}).  Although this model is very attractive due to its simplicity, one has to be careful when applying it (see \cite{PLOS}).

\section{Quantum instruments}
\label{QI}

We briefly present quantum instruments (see  also \cite{ENTROPY}  and 
basic papers \cite{DL70}-\cite{OO}; see Appendix \ref{apovm} for coupling with generalized observables given by POVMs.)  

\subsection{Quantum states and observables}

In quantum theory, it is postulated that every quantum system $S$ corresponds to a complex 
Hilbert space $\cH;$ denote the scalar product of two vectors by the symbol $\langle \psi_1\vert \psi_2\rangle.$
Throughout the present paper, we assume $\cH$ is {\it finite dimensional.}
States of the quantum system $S$ are represented by density operators acting in $\cH $ 
(positive semi-definite operators with unit trace).  Denote this state space by the symbol  $\bf{S}(\cH)$.

In quantum physics (especially quantum information theory), there are widely used notation conventions invented by Dirac:
a vector belonging to $\cH $ is symbolically denoted  as $\vert \psi\rangle;$ orthogonal projector on this vector is denoted 
as $\ketbra{\psi},$ it acts to the vector $\vert \xi \rangle$ as  $ \langle\psi\vert \xi\rangle \vert \psi\rangle.$     
  
Any density operator $\rho$ of rank one is of the form $\rho=\ketbra{\psi}$ with a unit-norm vector $\vert \psi\rangle.$
 In this case, $\vert \psi\rangle$ is called a {\em state vector}, so $\vert \psi\rangle \in \cH, \Vert \vert \psi\rangle \Vert= \sqrt{\langle \psi\vert \psi\rangle}=1.$

Observables (or physical quantities) of the quantum system $S$
are  represented by self-adjoint operators in $\cH.$   
Each observable $ A$ can be represented as 
\begin{equation}
\label{HH}
 A = \sum_x x E^{A}(x),
\end{equation}
where $E^{A}(x)$ is the spectral projection of the observable $A$ corresponding to 
 $x$ if $x$ is an eigenvalue of $A$; otherwise, let $E^{A}(x)=0$. 
We note that spectral projectors sum up to the unit operator $I:$
\begin{equation}
\label{HHar}
 \sum_x  E^{A}(x)=I.
\end{equation}
 We also remark that each orthogonal projector $E$ is self-adjoint and idempotent, i.e., $E^\star = E$ and $E^2=E.$ 

In what follows, we identify the states of the system with density operators and the observables of the system
with self-adjoint operators.

According to the standard interpretation of quantum mechanics, 
we can measure every observable $A$ in any state $\rho$ in principle, and then
we obtain one of its eigenvalues as the outcome of the measurement,
while quantum mechanics cannot in general predict the precise outcome, but only predict 
its probability distribution $\Pr\{A=x\|\rho\}$ by the {\it Born rule} as
\deq{\label{eq:Born}
\Pr\{A=x\|\rho\}=\Tr[E^{A}(x)\rho].
}
For a state vector $\vert \psi\rangle$, this leads to the relation 
\deq{\Pr\{A=x\|\vert \psi\rangle\}=\|E^{A}(x)\vert \psi\rangle\|^2,}
as $\Tr[E^{A}(x)\ketbra{\psi}]=\|E^{A}(x)\vert \psi\rangle\|^2$.

Thus, quantum mechanics determines the probability distribution of a single observable in a given state.
We now consider the problem of determination of  the joint probabilities of several observables determined.
There are two types of joint probabilities in quantum mechanics: one for simultaneous measurement and
the other for successive measurement.  We shall discuss them for two observables.

In quantum mechanics, two observables $A$ and $B$ are simultaneously measurable in any state $\rho$
if and only if $A$ and $B$ are commuting, i.e., $[A,B]=AB-BA=0$, and in this case, the joint probability
distribution of outcomes of simultaneous measurements of $A$ and $B$ in a state $\rho$ is give by
\deq{\label{eq:SM-1}
\Pr\{A=x,B=y\|\rho\}=\Tr[E^{A}(x)E^{B}(y)\rho] .}
For a state vector $\vert \psi\rangle$, this leads to the relation 
\deq{\label{eq:SM-2}\Pr\{A=x,B=y\|\rho\}=\|E^{A}(x)E^{B}(y)\vert \psi\rangle\|^2.}
\sloppy
Since $A$ and $B$ are commuting, in the above expressions we have $E^{A}(x)E^{B}(y)=E^{B}(y)E^{A}(x)$ and
the order of $A=x$ and $B=y$ are interchangeable.
Note that \Eq{SM-1} and \Eq{SM-2} can be obtained by the Born rule without a new postulate \cite{VN}; 
see Appendix  \ref{se:SM}.

How does quantum mechanics determine the joint probability distribution of successive
measurements?
Does the Born rule determine the joint probability of the outcomes of the $A$ measurement and the
subsequent $B$ measurement?
Suppose that the observable $A$ is measured in a state $\rho$ and the result $A=x$ is obtained and subsequently
the observable $B$ is measured and the result $B=y$ is obtained.  To determine the joint probability distribution
 $\Pr\{A=x,B=y\|\rho\}$ of their outcomes, we consider the conditional probability distribution of $B$
 given $A=x$ defined by
 \deq{
 \Pr\{A=x,B=y\|\rho\} = \Pr\{A=x\|\rho\} \Pr\{B=y|A=x\|\rho\}.
 }
The definition of conditional probability is based on the following reasoning.
The state of the system just after the $A$ measurement is considered to depend on the initial state $\rho$ and
the outcome $A=x$ so that we denote it by $\rho_{\{A=x\}}$.  Since the $B$ measurement is carried out
in this state, the Born rule applies to this state to obtain
 \deq{
 \Pr\{B=y|A=x\|\rho\}=\Pr\{B=y\|\rho_{\{A=x\}}\},
 }
 and hence the joint probability distribution is given by
 \deq{
\Pr\{A=x,B=y\|\rho\}
&=\Pr\{B=y\|\rho_{\{A=x\}}\}\Pr\{A=x\|\rho\}\nn\\
&=\Tr[E^B(y)\rho_{\{A=x\}}]\Tr[E^A(x)\rho].}

The Born rule gives the probability of the output in a given state.
But, in order to determine the joint probability distribution $\Pr\{A=x,B=y\|\rho\}$
we need another postulate to determine the state $\rho_{\{A=x\}}$ after the measurement.

\subsection{Von Neumann-L\"uders instruments}
\label{VNL}

In the conventional QMT, the {\em projection
postulate}\footnote{For observables given by self-adjoint operators with non-degenerate spectra, this postulate was suggested 
by von Neumann \cite{VN}. Then L\"uders extended it even to observables with degenerate spectra \cite{Luders}. 
For the latter, von Neumann used a more general rule (in the spirit of theory of quantum instruments).} is posited, stating:
in the measurement of an observable $A$, the input state $\rho$ is changed to the output state
\deq{\label{eq:projection}
\rho_{\{A=x\}}=\frac{E^{A}(x)\rho E^{A}(x)}{\Tr[E^{A}(x)\rho]}
}
provided that the measurement leads to the outcome $A=x$.\footnote{
This rule was adopted by Wang and Busemeyer \cite{WB13}.}
If the input state is the state vector $\vert \psi\rangle$, the output state is also the state vector 
$\vert \psi_{\{A=x\}}\rangle$
such that 
\deq{\vert \psi_{\{A=x\}}\rangle=\frac{E^{A}(x)\vert \psi\rangle}{\|E^{A}(x)\vert \psi\rangle\|}.
}
In this case, i.e., for the density operator $\rho=\ketbra{ \psi},$ we have the relations
\deq{
\rho_{\{A=x\}}=\frac{E^{A}(x)\rho E^{A}(x)}{\Tr[E^{A}(x)\rho]}
=\frac{E^{A}(x)\ketbra{\psi} E^{A}(x)}{\|E^{A}(x)\vert \psi\rangle\|^2}=\ketbra{\psi_{\{A=x\}}}.
}

Note that equality 
(\ref{HHar}) can be rewritten in the form:
\begin{equation}
\label{HHa}
 \sum_x \, E^{A}(x)^\star E^{A}(x) =I,
\end{equation}
and the Born rule is written as
\deq{\label{eq:Born2}
\Pr\{A=x\|\rho\}=\Tr[E^{A}(x)\rho E^{A}(x)].
}

According to the projection postulate, if observables $A$ and $B$ are successively measured
in this order in the initial input state $\rho$, the joint probability distribution of their outcomes is given by
\deq{
\Pr\{A=x,B=y\|\rho\}=\Tr[E^{B}(y)E^{A}(x)\rho E^{A}(x)].
}
For the state vector $\vert \psi\rangle$ we have
\deq{
\Pr\{A=x,B=y\|\vert \psi\rangle\}=\| E^{B}(y)E^{A}(x)\vert \psi\rangle\|^2.
}

Thus, in the conventional QMT the outcome probability distribution and
the state change caused by the measurement are uniquely determined by the observable $A.$  

However, the theory should also reflect the evident fact that the same quantum observable, say energy, can be measured 
with a variety of quantum measuring instruments. In the quantum formalism, these instruments are characterized by 
back-actions of measurements to systems' states. 
In modern QMT, a more flexible rule is adopted, in which the state change caused 
by the measurement is not uniquely determined by the observable to be measured.
There are many ways to measure the same observable.  

\subsection{The Davis-Lewis-Ozawa quantum instruments}

It has been known that the projection postulate is too restrictive to describe all the
physically realizable measurements of the observable $A$.  Davies-Lewis \cite{DL70} 
proposed to abandon this postulate and proposed a more flexible approach to QMT. 

The space $\cL(\cH)$ of linear operators in $\cH $ is a linear space over the complex numbers.\footnote{We recall that in this paper 
we use only finite dimensional Hilbert spaces. In the infinite dimensional case, one has to consider the space of all trace class operators.} Moreover,  $\cL(\cH)$ is the
complex Hilbert space with the scalar product, $\langle A\vert B\rangle= \rm{Tr} A^\star B.$ 
We  consider linear operators acting on it; they are called {\it superoperators}. 
The latter term is used to distinguish operators acting in the Hilbert spaces $\cH$ and $\cL(\cH).$
Otherwise superoperators are usual linear operators. In particular, for $\cT: \cL(\cH) \to \cL(\cH),$ 
there is well defined adjoint operator $\cT^\star: \cL(\cH) \to \cL(\cH).$  However, some basic notions  are specific for superoperators.   
A superoperator $\cT: \cL(\cH) \to \cL(\cH)$ is called {\it positive} if it maps the set 
of  positive semi-definite operators into itself. A superoperator is called {\it completely positive} if its natural extension 
$\cT\otimes \id$ to the tensor product $\cL(\cH)\otimes \cL(\cH)=\cL(\cH\otimes\cH)$ is again a positive superoperator
on $\cL(\cH)\otimes \cL(\cH)$.  (This notion is rather complicated technically. Therefore we shall not discuss it in more detail.) 

Consider now a general measurement on the system $S$.
The statistical properties of any measurement are characterized by 
\begin{itemize}
\item (i) the output probability 
distribution $\Pr\{\bx=x\| \rho\}$, the probability distribution of the output $\bx$ of the measurement
in the input state $\rho:$
\item (ii) the quantum state reduction $\rho\mapsto\rho_{\{\bx=x\}}$, the state change from 
the input state $\rho$ to the output state $\rho_{\{\bx=x\}}$ conditional upon the outcome 
$\bx=x$ of the measurement.
\end{itemize}
According to Davies--Lewis \cite{DL70} and Ozawa \cite{Oza84},
the modern quantum measurement theory postulates that 
any measurement of the system $S$ is described by a mathematical structure 
called a {\it quantum instrument}.
This is any map $x \to \cI(x)$, where for each real $x$, the map  $\cI(x)$ is a completely positive 
superoperator satisfying the normalization condition $\sum_x\Tr[\cI(x)\rho]=1$ for any
state $\rho.$ 

Davies--Lewis \cite{DL70} originally postulated that the superoperator
$ \cI(x)$ should be positive.  However, Yuen \cite{Yue87} pointed out that the Davies--Lewis
postulate is too general to exclude physically non-realizable instrument.  Ozawa \cite{Oza84} 
introduced complete positivity to ensure that every quantum instrument
is physically realizable.i
Thus, complete positivity is a sufficient condition for an instrument to be physically realizable.
On the other hand,  necessity is derived as follows \cite[p.~369]{O4}.
Every observable $A$ of a system $S$ is identified with the observable $A\otimes I$ of a
system $S+S'$ with any system $S'$ external to $S$\footnotemark.
\footnotetext{For example, the color $A$ of my eyes is not only the property
of my eyes $S$ but also the property $A\otimes I$ of myself $S+S'$.}
Then, 
every physically realizable instrument $I_A$ measuring $A$ should be identified
with the instrument $I_{A\otimes I}$ measuring $A\otimes I$
such that $I_{A\otimes I}(x)=I_A(x)\otimes \id$.  
This implies that $I_A(x)\otimes \id$ is again a positive superoperator,
so that $I_A(x)$ is completely positive.  Similarly, any physically realizable instrument
$\cI(x)$ measuring system $S$ should have its extended instrument $\cI(x)\otimes \id$
measuring system $S+S'$ for any external system $S'$.  This is fulfilled only
if $\cI(x)$ is completely positive.
Thus, complete positivity is a necessary condition for  $I_A$ to describe a physically 
realizable instrument.

Given a quantum instrument $\cI$, the output probability distribution for the input state $\rho$
is defined by the generalized Born rule in the trace-form, 
\begin{equation}
\label{eq:BRULEy}
\Pr\{\bx =x\|\rho\} := \Tr\; [\cI(x) \rho],
\end{equation}
and the quantum state reduction is defined by 
\begin{equation}
\label{eq:TRA4}
\rho \mapsto \rho_{\{\bx=x\}}:= \frac{\cI(x)\rho}{\Tr[ \cI(x)\rho]}.
\end{equation}

According to the Kraus theorem \cite{Kra71}, for any instrument $\cI$,
there exists a family $\{M_{xj}\}_{x,j}$ of operators, called the {\em measurement operators} for
$\cI$, in $\cH$ such that
\deq{\label{eq:MO-1}
\cI(x)\rho=\sum_{j}M_{xj}\rho M_{xj}^{\da}
} 
for any state $\rho$. 
In this case, we have
\deq{\label{eq:MO-2}
\sum_{xj}M_{xj}^{\da}M_{xj}=I.
}
Conversely, any family $\{M_{xj}\}_{x,j}$ of operators in $\cH$ satisfying \eq{MO-2} defines
an instrument $\cI$.

The above general formulation of quantum instruments reflects variety of real measuring 
instruments for the same system $S$ that measure an observable $A$ accurately or with some error.
A quantum instrument $\cI_A$ is called an instrument measuring an observable $A$,  or an $A$-measuring
instrument, if the output probability distribution satisfies Born's rule \eq{Born} for the $A$-measurement, i.e, 
\deq{\label{eq:BRULEya}
\Tr [\cI_A(x) \rho]=\Tr[E^{A}(x)\rho].
}
The projective $A$-measuring instrument is defined by
\deq{
\cI_A(x)\rho:=E^{A}(x)\rho E^{A}(x)
}
for any state $\rho$ and real number $x$.
Then, this instrument satisfies not only the Born formula \eq{Born} or \eq{Born2}, but also 
the projection postulate \eq{projection}.
Thus, the projection postulate is no longer the requirement for the measurement 
of the observable $A$ but only one type of the measurement of $A$.

Measurement operators $\{M_{xj}\}_{x,j}$ for
$\cI_A$ in $\cH$ satisfies the relation
\deq{\label{eq:MO-3}
E^{A}(x)=\sum_{j}M_{xj}^{\da}M_{xj}.
}
Conversely, any family $\{M_{xj}\}_{x,j}$ of operators in $\cH$ satisfying \eq{MO-3} defines
an $A$-measuring instrument $\cI_A$ by \Eq{MO-1}.
For the projective instrument $\cI_A$ measuring $A$ the measurement operators coincide
with the spectral projections, i.e., $M_{xj}=E^{A}(x)$ with $j\in\{1\}$.

\subsection{Quantum order effect from the quantum instrument viewpoint}

Finally, we offer a probabilistic comment on quantum instruments. Typically the main attention is given to Born's rule \eq{BRULEya}
as generating probabilities from quantum states, $\rho \mapsto \Pr\{A =x\|\rho \}.$ We would like to elevate the role of the state transform \eq{TRA4}, 
measurement back-action. It generates quantum probability conditioning. If after measurement of observable $A$ with outcome 
$A=x,$ one measures observable $B,$ then the probability to get output 
$B=y$ is given by Born's rule for state $\rho_{\{A=x\}}:$
\beq
\label{MA17}
\Pr\{B=y\vert A =x \| \rho\}= \Tr \;  [\cI_B(y)\rho_{\{A=x\}}]= \frac{\Tr  \; [\cI_B(y) \cI_A(x) \rho]}{\Tr  \;  [\cI_A(x)]}.
\eeq 
Now, by using quantum conditional probability we can define the sequential joint probability distribution of $A$ (first) and $B$ (last),
\beq
\label{MA17a}
\Pr\{A =x, B=y \| \rho\}=   \Pr\{A =x \| \rho\} \Pr\{B=y\vert A =x \| \rho\}= \Tr  \; [\cI_B(y) \cI_A(x) \rho].
\eeq 
It is clear that if superoperators $\cI_A(x)$ and $\cI_B(y)$ do not commute, i.e., 
\beq
\label{NC1}
[\cI_A(x), \cI_B(y)] =  \cI_A(x) \cI_B(y) -  \cI_B(y) \cI_A(x) \not=0,
\eeq 
 then generally 
$\Pr\{A =x, B=y \| \rho \} \not= \Pr\{B=y, A =x \| \rho \}.$ 
This is the mathematical representation of  QOE. 

In the framework of the quantum instrument theory, one has to distinguish noncommutativity of observables and noncommutativity of instruments. 
The standard noncommutativity condition
\beq
\label{NC2}
[A, B] \not =0,
\eeq 
describes incompatibility of observables, in the sense that their joint probability distribution (JPD), the JPD for their {\it simultaneous measurements}, is not well defined. 
So, condition (\ref{NC2}) is related to joint measurements, not to sequential measurement and not to QOE. The latter is characterized 
not via non-existence  of JPD, but via noncommutativity of the state updates, formalized via condition (\ref{NC1}).

Generally, noncommutativity of instruments (\ref{NC1}) does not imply noncommutativity of the observables to be measured 
(\ref{NC2}), whereas the converse statement holds as shown in the following theorem.

\bTheorem
For any $A$-measuring instrument $\cI_A$ and $B$-measuring instrument $\cI_B$,
if $[\cI_A(x),\cI_B(y)]=0$ for any $x,y$ then $[A,B]=0$.
\eTheorem
\bProof
\sloppy
It is well known that $\cI_A(x)^* X=T_A^*(X)E^{A}(x)$ and 
$\cI_B(y) Y=T_B^*(Y)E^{B}(y)$ for any $X,Y\in\cL(\cH)$, where $T_A^*=\sum_x\cI_A(x)^*$, 
$T_B^*=\sum_y\cI_B(y)^*$ and that $[T_A^*(X),E^{A}(x)]=0$ and $[T_B^*(Y),E^{B}(y)]=0$
for all $X,Y\in\cL(\cH)$ and $x,y$ \cite[Proposition 4.4]{Oza84}.
Suppose $ \cI_A(x) \cI_B(y) -  \cI_B(y) \cI_A(x) =0$ for every $x,y$.
Then, we have  $\cI_A(x)^* \cI_B(y)^*I= \cI_B(y)^* \cI_A(x)^*I$
and $\cI_A(x)^* E^{B}(y)= \cI_B(y)^* E^{A}(x)$.
Summing up all $y$ we have  $\cI_A(x)^* I= T_B^*(E^A(x))$ and $E^{A}(x)=T_B^*(E^A(x))$.
From $[T_B^*(Y),E^{B}(y)]=0$ we have $[E^{A}(x),E^{B}(y)]=0$ for all $x,y$.
Thus, $[A,B]=0$.
\eProof
We shall in fact construct instruments $\cI_A$ and $\cI_B$ showing the question order effect with
commuting observables $A$ and $B$, while the commutativity of $A$ and $B$ is necessary for
$\cI_A$ and $\cI_B$ to show the response replicability effect.

\subsection{Quantum instruments as representation of indirect measurements}

The basic model for construction of quantum instruments is based on  the scheme of indirect measurements. 
This scheme formalizes the following situation. As was permanently emphasized by Bohr (one of the founders of 
quantum mechanics), the results of quantum measurements are generated in the process of interaction of a system $S$ 
with a measurement apparatus $M.$ This apparatus consists of a complex physical device interacting with $S$ and a pointer 
that shows the result of measurement, say spin up or spin down. An observer can see only outputs of the pointer and he 
associates these outputs with the values of the observable $A$ for the system $S.$ So, the observer approaches only 
the pointer, not the system by itself. Whether the outputs of the pointer can be associated with the ``intrinsic properties'' of
$S$  is one of the main problems  of quantum foundations, it is still a topic for foundational debates 
\cite{B1}-\cite{CP77}.  Thus, the indirect measurement scheme involves:
\begin{itemize}
\item the states of the systems $S$ and the apparatus $M;$    
\item the operator $U$ representing the interaction-dynamics for the system $S+M;$ 
\item the meter observable $M_A$  giving outputs of the pointer of the apparatus $M.$ 
\end{itemize}

We shall make the following remark on the operator $U$ of the interaction-dynamics. As all operations in the quantum mechanics, 
it is a linear operator. In the quantum formalism, dynamics of the state of an isolated system is described by  
the Schr\"odinger equation and its evolution operator is unitary. In the indirect measurement scheme, it is assume that (approximately) the compound system 
$S+M$ is isolated. Hence, its evolution operator $U$ is unitary. 

Formally, an {\em indirect measurement model}, introduced in \cite{Oza84} as a ``(general) measuring process'',
is a quadruple 
$$
(\cK, \sigma, U, M_A)
$$ 
consisting of a Hilbert space $\cK,$   a density operator $\sigma \in {\bf S}(\cK),$ a unitary operator $U$ 
on the tensor product of the state spaces of $S$ and $M,$ $U: \cH\otimes\cK \to \cH\otimes\cK,$ and a self-adjoint operator $M_A$ on $\cK$.  
By this measurement model,  the Hilbert space $\cK$ describes the states of the apparatus  $M$, the unitary operator  $U$ 
describes the time-evolution of the composite system $S+ M$, the density operator $\sigma$ describes the initial state of the apparatus $M$, 
and the self-adjoint operator $M_A$ describes the meter observable of the apparatus $M.$ 
Then, the output probability distribution $\Pr\{A =x\|\rho\}$ in the system state $\rho \in \bf{S}(\cH)$
is given by
\beq
\label{MA1}
\Pr\{A =x\|\rho\}=\Tr[(I\otimes E^{M_A}(x))U(\rho\otimes\sigma)U^{\star}],
\eeq 
where $E^{M_A}(x)$ is  the spectral projection of $M_A$ for the eigenvalue $x.$

The change of the state $\rho$ of the system $S$  caused by the measurement for the outcome $A =x$ 
is represented with the aid of the map $\cI_A(x)$ in the space of density operators defined as 
\beq
\label{MA1A}
\cI_A(x)\rho =\Tr_{\cK}[(I\otimes E^{M_A}(x))U(\rho\otimes\sigma)U^{\star}],
\eeq 
where $\Tr_{\cK}$ is the partial trace over $\cK.$ 
Then,  the map  $x \mapsto \cI_A(x)$ turns out to be a quantum instrument. 
Thus, the statistical properties of the measurement realized by any indirect measurement 
model $(\cK,\sigma,U,M_A)$ is described by a quantum measurement.
We remark that conversely any quantum instrument can be represented via the indirect 
measurement model \cite{Oza84}.   Thus, quantum instruments mathematically 
characterize the statistical properties of all the physically realizable quantum 
measurements\footnotemark.

\footnotetext{It is an interesting problem whether instruments on a more general state space 
are realizable by an indirect measurement model $(\cK,\sigma,U,M_A)$.
This problem was studied extensively by  Okamura and Ozawa \cite{OO}.
Any discrete classical instrument is realizable.
There exists a non-realizable continuous classical instrument.
There also exists a non-realizable instrument for a local system in algebraic quantum field theory,
but all instruments are realizable within arbitrary error.
 A necessary and sufficient condition for any instrument on a von Neumann algebra 
 (which includes classical and quantum mechanics, and algebraic quantum field theory) to be precisely realizable 
by an indirect quantum measurement model
is known as the normal extension property (NEP) \cite{OO}.}

Now, we point to a few details which were omitted in the above considerations.  
The measuring interaction  between the system $S$ and the apparatus $M$ turns on at time $t_0,$ the time of measurement, 
and turns off at time $t=t_0 + \Delta t.$  We assume that the system $S$ and the apparatus $M$ do not interact each
other before $t_0$ nor after $t=t_0 + \Delta t$ and that the compound
system $S + M$ is isolated in the time interval
$(t_0, t).$ The {\it probe system} $P$ is defined to be the minimal
part of apparatus $M$ such that the compound system
$S+P$ is isolated in the time interval $(t_0, t).$ Then the above scheme is applied to the probe system $P,$ instead 
of the whole apparatus $M.$ The rest of the apparatus $M$ performs 
the pointer measurement on the probe $P.$ In particular, the unitary evolution operator $U$ describing the state-evolution of the system 
$S+P$ has the form $H= e^{-i\Delta t  H},$ where $H =H_S+ H_P+ H_{SP}$ is Hamiltonian of $S+P$ with the terms $H_S$ and $H_P$ representing 
the internal dynamics in the subsystems $S$ and $P$ of the compound system and $H_{SP}$ describing the interaction between the subsystems.   

Introduction of probe systems may be seen as unnecessary complication of the scheme of indirect measurements. 
However, it is useful if the apparatus $M$ is a very complex system that interacts (often in parallel) with many systems 
$S_j, j=1,2,...., m.$ Its different probes are involved  solely in interaction with the concrete systems, $P_j$ with $S_j.$ 
And the system $S_j+P_j$ can be considered as
an isolated system; in particular, from interactions with other systems $S_i$ and probes $P_i.$ 

The indirect measurement scheme is part of the theory of open quantum systems \cite{OQS}. 
Instead of a measurement apparatus $M,$ we can consider the surrounding environment $\bf{E}$ of the system 
$S$ (see \cite{AS1}-\cite{AS3} for applications to psychology).      

\section{Constructions of Instruments}

\subsection{Observables $A$ and $B$}

In modeling successive question-response experiments, such as the Clinton-Gore experiment, 
we consider two questions $A$ and $B$ for a subject to answer ``yes'' ($y$) or ``no'' ($n$)
and consider the joint probability $p(AaBb)$ for obtaining the answer $a$ ($a=y$ or $n$) 
for the question $A$ and the answer $b$  ($b=y$ or $n$) for the question $B$, if the question $B$ is
asked after the question $A$, and the analogous joint probability  $p(BbAa)$
 if the question $A$ is asked after the question $B$.

We models the above joint probability distributions by the joint probability distributions
of outcomes of successive measurements of 2-valued observables $A$ and $B$ in a quantum system
in a given state $\rho$ in the $A$--$B$ order and in the $B$--$A$ order.
According to the general QMT,  the joint probability distributions are well defined by
two 2-valued observables (represented by projections) $A$, $B$ in a fixed Hilbert space $\cH$,
a quantum state $\rho$, and an $A$-measuring instrument $\cI_A$ and a $B$-measuring instrument
$\cI_B$ by the relations
\deqed{
p(AaBb)=\Pr\{A=a,B=b\|\rho\}=\Tr[\cI_B(b)\cI_A(a)\rho],\\
p(BbAa)=\Pr\{B=b,A=a\|\rho\}=\Tr[\cI_A(a)\cI_B(b)\rho].
}

Let $A$ and $B$ be projections on a Hilbert space $\cH$.
They correspond to questions labeled $A$ and $B,$ respectively.
The eigenvalue 1 means the answer ``yes'' to the questions, 
and the eigenvalue 0 means the answer ``no'' to the questions.

Let $\C^{2}=\{\ket{0},\ket{1}\}^{\perp\perp}$ 
and $\C^{3}=\{\ket{0},\ket{1},\ket{2}\}^{\perp\perp}$, 
where ${}^\perp$ stands for the orthogonal complement in $\cH$,
so that $S^{\perp\perp}$ stands for the subspace spanned by a subset $S$ of $\cH$.
We suppose that  $\cH$ consists of three components 
$\cH=\cH_1\otimes\cH_2\otimes \cH_3$ such that $\cH_1=\cH_2=\C^{2}$ and $\cH_3=\C^{3}$.
The space $\cH$ is called the space of {\em mind states}, the first component 
$\cH_1$ is called the space of {\em belief states for question $A$}, the secant component
$\cH_2$ is called the space of {\em belief states for question $B$}, the third component
is called the space of {\em personality states}.

Denote by $I_1$,  $I_2$, and $I_3$ the identity operators on the first,  second,
and the third component of $\cH$,
respectively.  We let $A=\ketbra{1}\otimes I_2\otimes I_3$, and  
$B=I_1\otimes \ketbra{1}\otimes I_3$.
Thus, we consider the case where projections $A$ and $B$ commute.

\subsection{Instrument $\cI_A$ measuring $A$}\label{se:I_A}
We construct an instrument $\cI_A$ measuring $A$ as follows.
The instrument $\cI_A$ carries out a measurement of $A$ 
by a measuring interaction between the object $S$ described by the state space $\cH$
and the probe $P$ described by the state space 
$\cK=\C^{2}\otimes\C^{2}$, which is prepared in the state $\ket{00}$
just before the measuring interaction.
Denote by $I_4$ and $I_5$ the identity operators on the first and second component of $\cK$.
For the composite system $S+P,$ its state transformation 
during the measuring interaction is described by a unitary operator $U_A$ on $\cH\otimes\cK$
satisfying
\begin{align}
U_A:\ket{000}\ket{00}&\mapsto \ket{000}\ket{00},\label{eq:U_A-1}\\
U_A: \ket{010}\ket{00}&\mapsto \ket{010}\ket{01},\\
U_A: \ket{100}\ket{00}&\mapsto \ket{100}\ket{10},\\
U_A: \ket{110}\ket{00}&\mapsto \ket{110}\ket{11}\label{eq:U_A-4},\\
U_A:\ket{001}\ket{00}&\mapsto \ket{011}\ket{00},\label{eq:U_A-5}\\
U_A: \ket{011}\ket{00}&\mapsto \ket{011}\ket{01},\\
U_A: \ket{101}\ket{00}&\mapsto \ket{101}\ket{10},\\
U_A: \ket{111}\ket{00}&\mapsto \ket{101}\ket{11}\label{eq:U_A-8},\\
U_A:\ket{002}\ket{00}&\mapsto \ket{002}\ket{00},\label{eq:U_A-9}\\
U_A: \ket{012}\ket{00}&\mapsto \ket{002}\ket{01},\\
U_A: \ket{102}\ket{00}&\mapsto \ket{112}\ket{10},\\
U_A: \ket{112}\ket{00}&\mapsto \ket{112}\ket{11}\label{eq:U_A-12},
\end{align}
and the outcome of the measurement is obtained by measuring the meter observable 
\beq
M_A=\ketbra{1}\otimes I_5
\eeq
of the probe $P$.
Note that both $A$ and $M_A$ have the same spectrum $\{0,1\}$ and they are
projections.
We denote by $I_{\cH}$ and $I_{\cK}$ the identity operators on $\cH$ and $\cK$,
respectively, and denote by $M_A^\perp$ the orthogonal complement of the projection $M_A$,
 i.e., $M_A^\perp=I_{\cK}-M_A$.
 
Equations \eq{U_A-1}--\eq{U_A-4} describe the change of the mind state + the probe state
with the personality state $\ket{0}$.  In this case, the belief state is copied to the probe state
and the belief state does not change.  Thus, in the decision stage, measuring the first state of 
the probe state results in the decision equal to the belief for question A just before the
question-response process.

Equations \eq{U_A-5}--\eq{U_A-8} describe the change of the mind state + the probe state
with the personality state $\ket{1}$.  In this case, the belief state is copied to the probe state
as well but the belief state for question B changes if the belief states for question A and 
question B are equal.  The decision stage results in the preset belief for question A as well.

Equations \eq{U_A-9}--\eq{U_A-12} describe the change of the mind state + the probe state
with the personality state $\ket{2}$.  In this case, the belief state is copied to the probe state
as well but the belief state for question B changes if the belief states for question A and 
question B are not equal.  The decision stage results in the preset belief for question A as well.

\sloppy
Suppose that the probe $P$  is prepared in the state $\ket{00}$ just before the measuring
interaction, the measuring process described by the indirect measurement model
$(\cK,\ket{00},U_A,M_A)$ defines
the instrument $\cI_A$ by
\begin{align}\label{eq:ins-X}
\cI_A(a)\rh
&=\Tr_{\cK}\left[\left(I_{\cH}\otimes P^{M_A}(a)\right) U_A(\rho\otimes \ketbra{00})U_A^{\da}
\left(I_{\cH}\otimes P^{M_A}(a)\right)\right],
\end{align}
for any density operator $\rh$ on $\cH$,
where $\Tr_{\cK}$ stands for the partial trace over $\cK$ and 
$P^{M_A}(a)$ denotes the spectral projection of an observable $M_A$ for
 $a$, i.e., $P^{M_A}(0)=M_A^\perp$ and $P^{M_A}(1)=M_A$.
 Consequently, 
\deqed{
\cI_A(0)\rh
&=\Tr_{\cK}[(I_{\cH}\otimes M_A^\perp) U_A(\rho\otimes \ketbra{00})U_A^{\da}
(I_{\cH}\otimes M_A^\perp)
],\\
\cI_A(1)\rh
&=\Tr_{\cK}[(I_{\cH}\otimes M_A)U_A(\rho\otimes\ketbra{00})U_A^{\da}
(I_{\cH}\otimes M_A)
]
}
for any $\rho$.

The instrument $\cI_A$ determines 
the probability distribution $\Pr\{\ba=a\|\rho\}$ of the outcome $\ba$ of the measurement,
where $a=0,1$, and 
the state change $\rh\mapsto\rh_{\{\ba=a\}}$ caused by the measurement is determined by
\begin{align}
\Pr\{\ba=a\|\rho\}&=\Tr[\cI_A(a)\rh],\\
\rh\mapsto \rh_{\{\ba=a\}}&=\frac{\cI_A(a)\rh}{\Tr[\cI_A(a)\rh]}.
\end{align}

From  Appendix \ref{ap:1},
it follows that the instrument $\cI_A$ measures the observable $A$,
i.e., 
\beql{I_A-A}
\begin{aligned}
\Pr\{\ba=0\|\rho\}&=\Tr[\cI_A(0)\rh]=\Tr[A^\perp\rh],\\
\Pr\{\ba=1\|\rho\}&=\Tr[\cI_A(1)\rh]=\Tr[A\rh]
\end{aligned}
\eeq
for any density operator $\rh$ on $\cH$, and we obtain 
\begin{align}\label{eq:IA}
\lefteqn{\cI_A(a)\rh=
\sum_{\be}\left(\ket{a,\be,0}\av{a,\be,0|\rho|a,\be,0}\bra{a,\be,0}\right.}\quad\nn\\
&\left.+\ket{a,a\Perp,1}\av{a,\be,1|\rho|a,\be,1}\bra{a,a\Perp,1}+
 \ket{a,a,2}\av{a,\be,2|\rho|a,\be,2}\bra{a,a,2}\right).
\end{align} 
for any density operator $\rho$ on $\cH$ and
$a=0,1$ as a general form of the instrument $\cI_A$ obtained by the indirect measurement
model $(\cK,\ket{00},U_A,M_A)$.

\subsection{Instrument $\cI_B$ measuring $B$}
The instrument $\cI_B$ is constructed with the same probe system  $P$ 
 prepared in the state $\ket{00}$ in an analogous manner with the instrument $\cI_A$.
The unitary operator $U_B$ on $\cH\otimes\cK$, describing the time evolution
of the composite system $S+P$ during the measuring interaction,
is supposed to satisfy
\begin{align}
U_B:\ket{000}\ket{00}&\mapsto \ket{000}\ket{00},\label{eq:U_B-1}\\
U_B: \ket{010}\ket{00}&\mapsto \ket{010}\ket{01},\\
U_B: \ket{100}\ket{00}&\mapsto \ket{100}\ket{10},\\
U_B: \ket{110}\ket{00}&\mapsto \ket{110}\ket{11}\label{eq:U_B-4},\\
U_B:\ket{001}\ket{00}&\mapsto \ket{101}\ket{00},\label{eq:U_B-5}\\
U_B: \ket{011}\ket{00}&\mapsto \ket{011}\ket{01},\\
U_B: \ket{101}\ket{00}&\mapsto \ket{101}\ket{10},\\
U_B: \ket{111}\ket{00}&\mapsto \ket{011}\ket{11}\label{eq:U_B-8},\\
U_B:\ket{002}\ket{00}&\mapsto \ket{002}\ket{00},\label{eq:U_B-9}\\
U_B: \ket{012}\ket{00}&\mapsto \ket{112}\ket{01},\\
U_B: \ket{102}\ket{00}&\mapsto \ket{002}\ket{10},\\
U_B: \ket{112}\ket{00}&\mapsto \ket{112}\ket{11}\label{eq:U_B-12},
\end{align}
and the meter observable on $\cK$ is given by
\beq
M_B=I_4\otimes \ketbra{1}.
\eeq
Then the instrument $\cI_B$ of this measuring process $(\cK,\ket{00},U_B,M_B)$ 
is defined by
\begin{align}\label{eq:ins-B1}
\cI_B(b)\rh
&=\Tr_{\cK}\left[\left(I_{\cH}\otimes P^{M_B}(b)\Big)U_B(\rho\otimes \ketbra{00})
U_B^{\da}\Big(I_{\cH}\otimes P^{M_B}(b)\right)\right]
\end{align}
for any density operator $\rh$ on $\cH$ and $b=0,1$.
Then we have
\begin{align}\label{eq:IB}
\cI_B(b)\rh&=
\sum_{\al}\ket{\al,b,0}\av{\al,b,0|\rho|\al,b,0}\bra{\al,b,0}\nn\\
&+\ket{b\Perp,b,1}\av{\al,b,1|\rho|\al,b,1}\bra{b\Perp,b,1}+
 \ket{b,b,2}\av{\al,b,2|\rho|\al,b,2}\bra{b,b,2}.
\end{align} 
for any density operator $\rho$ on $\cH$ and for any $b=0,1$.
Thus, the probability distribution of 
 the outcome $\bb$ of 
the instrument $\cI_B$ is given by
\deqed{\label{eq:I_B-A}
\Pr\{\bb=0\|\rho\}&=\Tr[\cI_B(0)\rh]=\Tr[B^\perp\rh],\\
\Pr\{\bb=1\|\rho\}&=\Tr[\cI_B(1)\rh]=\Tr[B\rh]
}
for any density operator $\rh$ on $\cH$, and hence the instrument $\cI_B$
measures the observable $B$.

\section{Postulates for quantum models}

Our model presented above satisfies the following postulates.

{\em Postulate 1: A subject's mind state is represented by a state vector in or
a density operator on a multidimensional feature space (technically, an $N$-dimensional Hilbert space).}

The above postulate extends Postulate 1 in Wang-Busemeyer \cite{WB13} to include density operators.
State vectors $\ket{\psi_j}$ can be mixed in two way. For complex numbers $\al_j$ with $\sum_j\, |\al_j|^2=1$,
we can make a new state vector, called the superposition, $\sum_{j}\,\al_j\ket{\psi_j}$,  and alternatively
 for positive numbers $p_j$ with $\sum_j\,  p_j=1$, we can make a new state, called the mixture, 
represented by the density operator  $\sum_{j}\, p_j\ketbra{\psi_j}$. 
We use density operators if the state is a probability mixture (not a superposition) of state vectors. 

{\em Postulate 2: A potential response to  a question is represented by a subspace of the multidimensional
feature space.}

This repeats Postulate 2 in  \cite{WB13}.
In our model, we consider two questions A and B corresponding to the two subspaces $\cA$ and $\cB$,
which are represented by the two projections $A$ and $B$ that project the whole space $\cH$ onto $\cA$ 
and $\cB$, respectively.

 {\em Postulate 3: The probability of responding to an opinion question equals the squared length 
 of the projection of the state vector onto the response subspace, or equals the trace of the projection
 of the density operator onto the response subspace. }

This extends Postulate 3 in \cite{WB13} to include density operators.
According to \Eq{I_A-A},
in our model the probability that a subject in a state vector $\kps$ responds ``yes'' or ``no'' to the opinion question A 
is given by
\deqed{ 
\Pr\{A=y\|\kps\}&=\|A\kps\|^2,\\
\Pr\{A=n\|\kps\}&=\|A^{\perp}\kps\|^2.
}
From \Eq{I_B-A} analogous relations hold for question B.
According to \Eq{I_A-A},
for the ensemble of subjects represented by a density operator $\rho$ the frequency of responses  
``yes'' or ``no'' to the opinion question A  is given by
\deqed{ 
\Pr\{A=y\|\rho\}&=\Tr[A\rho]=\Tr[A\rho A],\\
\Pr\{A=n\|\rho\}&=\Tr[A\p \rho]=\Tr[A\p\rho A\p].
}
From \Eq{I_B-A} analogous relations hold for question B.

{\em Postulate 4: The updated mind state after deciding an answer to a question is determined by the instrument 
corresponding to the question.}

This postulate is more general than Postulate 4 in \cite{WB13}.
According to \Eq{MO-1} and  \Eq{MO-2} for any $A$-measuring instrument $\cI_A$, there exists a family $\{M_{1j}, M_{0j}\}_{j}$
 of operators such that 
\deq{
\cI_A(1)\rho&=\sum_{j}M_{1j}\rho M_{1j}^{\da},\\
A&=\sum_{j}M_{1j}^{\da}M_{1j},\label{eq:A-1}\\
\cI_A(0)\rho&=\sum_{j}M_{0j}\rho M_{0j}^{\da},\\
A^{\perp}&=\sum_{j}M_{0j}^{\da}M_{0j}.\label{eq:A-2}
}
If the state before answering the question $A$ is $\rho=\ketbra{\psi}$, the updated mind state $\rho_{\{A=y\}}$
after deciding an answer $A=y$ to the question $A$ is determined by the instrument $\cI_A$ as
\deqed{\rho_{\{A=y\}}&=\frac{\cI_A(1)\rho}{\Tr[\cI_A(1)\rho]}=\frac{\sum_{j}M_{1j}\ketbra{\psi}M_{1j}^{\da}}
{\|A\kps\|^2},\\
\rho_{\{A=n\}}&=\frac{\cI_A(0)\rho}{\Tr[\cI_A(0)\rho]}=\frac{\sum_{j}M_{0j}\ketbra{\psi}M_{0j}^{\da}}
{\|A^{\perp}\kps\|^2}.}
According to \Eq{IA}, an explicit form of the family $\{M_{1j}, M_{0j}\}_{j}$ is given with parameter $j=(\be,\ga)$
as follows.
\deqed{
M_{1,(\be,0)}=\ketbra{1,\be,0},\
M_{1,(\be,1)}=\ket{1,0,1}\bra{1,\be,1},\
M_{1,(\be,2)}= \ket{1,1,2}\bra{1,\be,2}.\\
M_{0,(\be,0)}=\ketbra{0,\be,0},\
M_{0,(\be,1)}=\ket{0,1,1}\bra{0,\be,1},\
M_{0,(\be,2)}= \ket{0,0,2}\bra{0,\be,2}.}
Thus, we reproduce Eqs.~\eq{A-1} and \eq{A-2}:
\deqs{
\sum_{\be.\ga}M_{1,(\be,\ga)}^*M_{1,(\be,\ga)}&=\sum_{\be,\ga} \ketbra{1,\be,\ga}=\ketbra{1}\otimes I_2\otimes I_3=A,\\
\sum_{\be.\ga}M_{0,(\be,\ga)}^*M_{0,(\be,\ga)}&=\sum_{\be,\ga} \ketbra{0,\be,\ga}=\ketbra{0}\otimes I_2\otimes I_3=A\p.}

Wang-Busemeyer \cite{WB13} posed 

{\em Postulate 4 (Wang-Busemeyer} \cite{WB13}{\em ): The updated belief state after deciding an answer 
to a question equals the normalized projection on the subspace representing the answer. }

This condition corresponds to the projective $A$-measuring instrument $\cI_A^{proj}$ such that
$\cI_A^{proj}(1)\rho=A\rho A$ and $\cI_A^{proj}(0)\rho=A\p\rho A\p$.  
This instrument has the measurement operators $\{A,A\p\}$
so that 
\deqed{
\rho_{\{A=y\}}&=\frac{\cI_A(1)\rho}{\Tr[\cI_A(1)\rho]}=\frac{A\ketbra{\psi}A}{\|A\kps\|^2},\\
\rho_{\{A=n\}}&=\frac{\cI_A(0)\rho}{\Tr[\cI_A(0)\rho]}=\frac{A\p\ketbra{\psi}A\p}{\|A^{\perp}\kps\|^2},
}
or states changes as
\deqed{
\kps\mapsto \ket{\psi_{\{A=1\}}}=\frac{A\kps}{\|A\kps\|},\\
\kps\mapsto \ket{\psi_{\{A=0\}}}=\frac{A\p\kps}{\|A\p\kps\|}.
}

Thus, our model satisfies Postulates 1-3 consistent with those posed by Wang-Busemeyer \cite{WB13}.
Moreover, our model generalizes Postulates 4 posed by Wang-Busemeyer \cite{WB13} equivalent
to the projection postulate, so that we allow any $A$-measuring instrument for question A,

Last but not least, we should pose an additional postulate on the ensemble of subjects, which is not
explicit but satisfied by the Wang-Busemeyer model \cite{WB13}.

{\em Postulate 5: If a sequence of question-response experiments is carried out on the same ensemble of
individual subjects, the joint probability of the sequence of responses is a probability mixture of 
those joint probabilities with respect to the frequency of individual mind states in the ensemble. 
}

If a sequence of question-response experiments is carried out on the same ensemble of
individual subjects, and a randomly chosen individual subject has the state $\ket{\psi_j}$ with probability
$p_j$, then we suppose that the joint probability of the sequence of responses 
$X_1=x_1,\ldots,X_n=x_n$,
where $X_k=A$ or $X_k=B$ and $x_k=0$ or $x_k=1$ for all $k=1,\ldots,n$, is given by
\deq{
\Pr\{X_1=x_1,\ldots,X_n=x_n\|\rho\}
=\Tr[\cI_{X_n}(x_n)\cdot\cI_{X_1}(x_1)\rho],
} 
where $\rho=\sum_{j}p_j\ketbra{\psi_j}$.
Then, we have
\deq{
\lefteqn{
\Pr\{X_1=x_1,\ldots,X_n=x_n\|\rho\}}\quad\nn\\
&\quad=
\sum_{j}p_j\Pr\{X_1=x_1,\ldots,X_n=x_n\|\ket{\psi_j}\}.
}
Thus, Postulate 5 is satisfied by our model. 

The structure of this postulate is discussed in more detail in Appendix \ref{apostulate5}.

\section{Mind states}

In modeling successive question-response experiments, such as the Clinton-Gore experiment, 
we assume that the subject's mind state is represented by the space  $\Om=\{0,1\}^2\times\{0,1,2\}$.
Here we suppose that any individual subject has one of the mind states, $\om\in\Om$, 
and any statistical ensemble of the subjects is characterized by a probability distribution 
on $\Om$.
Thus, the statistical data under consideration should be explained solely by one of the probability
distributions on $\Om$.
The mind state $\al$ represents the state in which the subject will answer 
``yes'' for the question A if  $\al=1$ and answer ``no'' otherwise.
The mind state $\be$ represents the analogous state for the question B.
Thus, the mind state $(\al,\be)$ represents the subject's prior belief as long as 
the questions A and B are concerned.
For $(\al,\be,\ga)\in\Om$ we write $\de_{(\al,\be,\ga)}=\ketbra{\al,\be,\ga}\in\cH$.
In this way, we identify the mind state $(\al,\be,\ga)\in\Om$ with the quantum
state $\ketbra{\al,\be,\ga}\in\cH$, and any probability distribution  $\mu$ of the
mind states $\om\in\Om$ with the quantum state 
$\hat{\mu}=\sum_{\om}\mu(\om)\de_{\om}$.
The dynamics of producing the answer to the question and preparing the mind
state for the next question is supposed to be described as a process of quantum 
measurement, or equivalently a mathematical object called a quantum instrument.
 
Our previously defined two quantum instruments $\cI_A$ and $\cI_B$ describe the
measurement on the mind state as follows.

\begin{Theorem}
For any $(\al,\be,\ga)\in\Om$ and $a,b=0,1$,
we have 
\begin{align}
\cI_A(a)\de_{(\al,\be,0)}&=\de_{\al}(a)\de_{(\al,\be,0)},\\
\cI_A(a)\de_{(\al,\be,1)}&=\de_{\al}(a)\de_{(\al,\al\Perp,1)},\\
\cI_A(a)\de_{(\al,\be,2)}&=\de_{\al}(a)\de_{(\al,\al,2)},\\
\cI_B(b)\de_{(\al,\be,0)}&=\de_{\be}(b)\de_{(\al,\be,0)},\\
\cI_B(b)\de_{(\al,\be,1)}&=\de_{\be}(b)\de_{(\be\Perp,\be,1)},\\
\cI_B(b)\de_{(\al,\be,2)}&=\de_{\be}(b)\de_{(\be,\be,2)}.
\end{align}
\end{Theorem}
\bProof
The relations follow from \Eq{IA} and \Eq{IB} by  routine computations.
\eProof

The mind state $\ga$ represents the personality of the subject in such a way that
if $\ga=1$, the subject changes his/her mind to prepare the answer for the other 
question to be the opposite to the previous answer, and if $\ga=2$, the subject changes
his/her mind to prepare the answer for the other question to be the same as the
previous answer; on the other hand, if $\ga=0$, the subject's mind is so robust 
that the question will not affect his/her mind.  

Thus, the state change is the Bayesian update if $\ga$ has only the value $\ga=0$, 
and yet if the value $\ga=1$ or $\ga=2$ is allowed, the statistics of the answers 
does not follows the Bayesian update rule.

The following considerations considerably simplify our manipulations of 
instruments $\cI_A$ and $\cI_B$ for arbitrary density operators $\rho$.

From Appendix \ref{ap:2}, for any density operator $\rh$ we have
\beql{Affine-IA}
\begin{aligned}
\cI_A(a)\rh&=\sum_{\al,\be,\ga}\mu(\al,\be,\ga)\cI_A(a)\de_{(\al,\be,\ga)}=\cI_A(a)\rh',
\\
\cI_B(b)\rh&=\sum_{\al,\be,\ga}\mu(\al,\be,\ga)\cI_B(b)\de_{(\al,\be,\ga)}=\cI_B(b)\rh',
\end{aligned}
\eeq
where $\mu(\al,\be,\ga)=\av{\al,\be,\ga|\rh|\al,\be,\ga}$, and 
\deq{\label{eq:Affine-IC}
\rh'=\sum_{\al,\be,\ga} \mu(\al,\be,\ga)\de_{(\al,\be,\ga)}.}

Therefore, the operations of $\cI_A$ and $\cI_B$ for arbitrary density operators $\rho$ are reduced to 
the operations for the density operators $\rho'$ of the form of \Eq{Affine-IC}, which are diagonal
in the $\ket{\al,\be,\ga}$ basis.

\section{Response Replicability Effect (RRE)}

Here, we consider two properties of successive measurements
of observables $A$ and $B$.

In A-A and A-B-A paradigms, the response to A is repeated (with probability 1). 
We call this response replicability effect (RRE).

In A-B vs B-A paradigm, the joint probabilities of the two responses are different on
a set of states with a positive Lebesgue measure. This is the question order effect (QOE).

The psychological problem raised by the Clinton-Gore experiment is as follows.  
We are given two 
observables 
$A$ and $B$ whose joint probability distributions (JPDs) 
of successive measurements shows QOE and is naturally considered to satisfy RRE.
We cannot represent the JPDs as the JPDs of classical random variables $A$ and $B$
as defined by Kolmogorov, 
which does not show QOE,  nor the JPDs of the outcomes of successive measurements of 
non-commuting quantum observables $A$ and $B$ 
as defined by von Neumann and L\"{u}ders, which does not satisfy RRE.

We shall show that instruments $\cI_A$ and $\cI_B$
have both RRE in this section,
and QOE in the next section.

\begin{Theorem}
The instruments $\cI_A$ and $\cI_B$ have the Response Replicability Effect;
namely, they satisfy the following relations.
\benum
\item $\sum_{a}\Tr[\cI_A(a)\cI_A(a)\rh]=1$ for any density operator $\rh$.
\item $\sum_{b}\Tr[\cI_B(b)\cI_B(b)\rh]=1$ for any density operator $\rh$.
\item $\sum_{a,b}\Tr[\cI_A(a)\cI_B(b)\cI_A(a)\rh]=1$ for any density operator $\rh$.
\item $\sum_{a,b}\Tr[\cI_B(b)\cI_A(a)\cI_B(b)\rh]=1$ for any density operator $\rh$.
\eenum
\end{Theorem}
\begin{Proof}
From Eq.~\eq{Affine-IA}  we assume without any loss
of generality that $\rho=\de_{(\al,\be,\ga)}$ for some $\al,\be,\ga$. 
For any $\al,\be$, we have
\begin{align*}
\sum_{a}\Tr[\cI_A(a)\cI_A(a)\de_{(\al,\be,0)}]
&=\sum_{a}\de_{\al}(a)\Tr[\cI_A(a)\de_{(\al,\be,0)}]\\
&=\sum_{a}\de_{\al}(a)\Tr[\de_{(\al,\be,0)}]\\
&=\Tr[\de_{(\al,\be,0)}]\\
&=1.
\end{align*}
Similarly, we have
\begin{align*}
\sum_{a}\Tr[\cI_A(a)\cI_A(a)\de_{(\al,\be,1)}]
&=1,\\
\sum_{a}\Tr[\cI_A(a)\cI_A(a)\de_{(\al,\be,2)}]
&=1.
\end{align*}
Thus, relation (i) follows.  Relation (ii) follows analogously.
For any $\al,\be$, we have
\begin{align*}
\sum_{a,b}\Tr[\cI_A(a)\cI_B(b)\cI_A(a)\de_{(\al,\be,0)}]
&=\sum_{a,b}\de_{\al}(a)\Tr[\cI_A(a)\cI_B(b)\de_{(\al,\be,0)}]\\
&=\sum_{a,b}\de_{\al}(a)\de_{\be}(b)\Tr[\cI_A(a)\de_{(\al,\be,0)}]\\
&=\sum_{a,b}\de_{\al}(a)\de_{\be}(b)\de_{\al}(a)\Tr[\de_{(\al,\be,0)}]\\
&=\Tr[\de_{(\al,\be,0)}]\\
&=1.
\end{align*}
Similarly,
\begin{align*}
\sum_{a,b}\Tr[\cI_A(a)\cI_B(b)\cI_A(a)\de_{(\al,\be,1)}]
&=1,\\
\sum_{a,b}\Tr[\cI_A(a)\cI_B(b)\cI_A(a)\de_{(\al,\be,2)}]
&=1.
\end{align*}
Thus, relation (iii) follows. Relation (iv) follows in a similar way.
\end{Proof}

\section{Question Order Effect (QOE)}

In what follows, 
we shall show that instruments $\cI_A$ and $\cI_B$
have the Question Order Effect (QOE).

Let  $\hat{\mu}=\sum_{\al,\be,\ga}\mu(\al,\be,\ga)\de_{(\al,\be,\ga)}$ for any probability
distribution $\mu$ on $\Om$, 
we write 
\begin{align}
p(AaBb)&=\Tr[\cI_B(b)\cI_A(a)\hat{\mu}],\\
p(BbAa)&=\Tr[\cI_A(a)\cI_B(b))\hat{\mu}],
\end{align}
where $a,b=0,1$.  We will write $Ay,An,By,Bn$ instead of $A1,A0, B1,B0$.
we have
\begin{align*}
\lefteqn{p(AaBb)}\qquad\\
&=\sum_{\al,\be,\ga}\mu(\al,\be,\ga)\Tr[\cI_B(b)\cI_A(a)\de_{(\al,\be,\ga)}]\\
&=
\sum_{\al,\be}\mu(\al,\be,0)\de_{\al}(a)\de_{\be}(b)+
\mu(\al,\be,1)\de_{\al}(a)\de_{\al\Perp}(b)
+\mu(\al,\be,2)\de_{\al}(a)\de_{\al}(b)\\
&=
\mu(a,b,0)
+\de_{a\Perp}(b)\sum_{\be}\mu(a,\be,1)
+\de_{a}(b)\sum_{\be}\mu(a,\be,2).
\end{align*}
Similarly,
\begin{align*}
p(BbAa)
&=\mu(a,b,0)
+\de_{b\Perp}(a)\sum_{\al}\mu(\al,b,1)
+\de_{b}(a)\sum_{\al}\mu(\al,b,2).
\end{align*}
Thus, 
\begin{align}
p(A{y}B{y})&=\mu(1,1,0)+\mu(1,1,2)+\mu(1,0,2).\label{eq:JPD-1}\\
p(A{y}B{n})&=\mu(1,0,0)+\mu(1,0,1)+\mu(1,1,1).\label{eq:JPD-2}\\
p(A{n}B{y})&=\mu(0,1,0)+\mu(0,1,1)+\mu(0,0,1).\label{eq:JPD-3}\\
p(A{n}B{n})&=\mu(0,0,0)+\mu(0,0,2)+\mu(0,1,2).\label{eq:JPD-4}\\
p(B{y}A{y})&=\mu(1,1,0)+\mu(1,1,2)+\mu(0,1,2).\label{eq:JPD-5}\\
p(B{n}A{y})&=\mu(1,0,0)+\mu(1,0,1)+\mu(0,0,1).\label{eq:JPD-6}\\
p(B{y}A{n})&=\mu(0,1,0)+\mu(0,1,1)+\mu(1,1,1).\label{eq:JPD-7}\\
p(B{n}A{n})&=\mu(0,0,0)+\mu(0,0,2)+\mu(1,0,2).\label{eq:JPD-8}
\end{align}

\begin{Theorem}\label{th:QOE}
The instruments $\cI_A$ and $\cI_B$ have the Question Order Effect;
namely, we have the following statements.
Let 
$\rho$ be a density operator on $\cH$.
\benum
\item 
 The relation
\beql{QOE-11}
 \Tr[\cI_B(1)\cI_A(1)\rho]
 =\Tr[\cI_A(1)\cI_B(1)\rho]
 \eeq
holds if and only if $\av{1,0,2|\rho|1,0,2}=\av{0,1,2|\rho|0,1,2}$.

\item 
The relation
\beql{QOE-10}
 \Tr[\cI_B(0)\cI_A(1)\rho]
 =\Tr[\cI_A(1)\cI_B(0)\rho]
 \eeq
holds if and only if  $\av{1,1,1|\rho|1,1,1}=\av{0,0,1|\rho|0,0,1}$.

\item 
 The relation
\beql{QOE-01}
 \Tr[\cI_B(1)\cI_A(0)\rho]
 =\Tr[\cI_A(0)\cI_B(1)\rho]
 \eeq
holds if and only if  $\av{1,1,1|\rho|1,1,1}=\av{0,0,1|\rho|0,0,1}$.

\item 
 The relation
\beql{QOE-00}
 \Tr[\cI_B(0)\cI_A(0)\rho]
 =\Tr[\cI_A(0)\cI_B(0)\rho]
 \eeq
holds if and only if $\av{1,0,2|\rho|1,0,2}=\av{0,1,2|\rho|0,1,2}$.

\item
One of Eqs.~\eq{QOE-11}, \eq{QOE-10}, \eq{QOE-01}, \eq{QOE-00}
for a general density operator $\rho$
holds only on a set of density operators  $\rho$ with Lebesgue measure 0.

\eenum
\end{Theorem}
\bProof
\Eq{QOE-11} holds if and only $p(AyBy)=p(ByAy)$.
Thus, assertion (i) follows from 
Eqs.~\eq{Affine-IA}, 
\eq{JPD-1}, and \eq{JPD-5}.
Assertions (ii)--(iv) follows similarly.
The last assertion follows from the fact that each relation holds
on a submanifold of the space of density operators 
with co-dimension 1.
\eProof

\section{QQ-equality}

If two questions, adjacent to each other, are asked in different orders, then the quantum model of measurement order 
makes an a priori and parameter-free prediction, named the QQ equality \cite{WSSB14,WB13,BB12}:
\begin{align}\label{eq:QQE}
q&=[p(ByAy)+p(BnAn)]-[p(AyBy)+p(AnBn)]\nn\\
&=[p(AyBn)+p(AnBy)]-[p(ByAn)+p(BnAy)] =0.
\end{align}
Note that the formula in \cite[P.~9435, left column]{WSSB14} is not correct
in that the first and second lines are not equal but have the opposite signs;
the correct definition here is due to \cite{WB13}.
We have
\begin{align}
p(AyBy)-p(ByAy)&=\mu(1,0,2)-\mu(0,1,2),\\
p(AnBn)-p(BnAn)&=\mu(0,1,2)-\mu(1,0,2),\\
p(AyBn)-p(BnAy)&=\mu(1,1,1)-\mu(0,0,1),\\
p(AnBy)-p(ByAn)&=\mu(0,0,1)-\mu(1,1,1).
\end{align}
Thus, 
it is easy to check that the QQ equality holds and we have

\begin{Theorem}\label{th:QQE}
The instruments $\cI_A$ and $\cI_B$ satisfy 
the QQ equality.
\end{Theorem}

\section{QQE-renormalizations}
We have shown that joint probabilities of responses to questions obtained from our model
satisfies QQE, so that if the original data did not satisfy QQE we could not reproduce those 
data from our model with arbitrary mind state $\rho$.  

Given the joint probabilities
$p(AyBy)$, $p(AyBn)$, $p(AnBy)$, $p(AnBn)$, $p(ByAy)$,
$p(ByAn)$, $p(BnAy)$, and $p(BnAn)$,
let 
\deq{
S_1&=\frac{p(AyBy)+p(AnBn)+p(ByAy)+p(BnAn)}{2},\\
S_2&=\frac{p(AyBn)+p(AnBy)+p(ByAn)+p(BnAy)}{2}.
}
Then, if QQE holds, we have
\deq{
p(AyBy)+p(AnBn)&=p(ByAy)+p(BnAn)=S_1,\\
p(AyBn)+p(AnBy)&=p(ByAn)+p(BnAy)=S_2.
}
Thus, in order for arbitrary $p(\cdots)$  to satisfy QQE
it is natural to ``renormalize'' them as $\q(\cdots)$ by 
\deq{
\q(AyBy)=
S_1\times
\frac{p(AyBy)}{p(AyBy)+p(AnBn)},\\
\q(AnBn)=
S_1\times
\frac{p(AnBn)}{p(AyBy)+p(AnBn)},\\
\q(ByAy)=
S_1\times
\frac{p(ByAy)}{p(ByAy)+p(BnAn)},\\
\q(BnAn)=
S_1\times
\frac{p(BnAn)}{p(ByAy)+p(BnAn)},\\
\q(AyBn)=
S_2\times
\frac{p(AyBn)}{p(AyBn)+p(AnBy)},\\
\q(AnBy)=
S_2\times
\frac{p(AnBy)}{p(AyBn)+p(AnBy)},\\
\q(ByAn)=
S_2\times
\frac{p(ByAn)}{p(ByAn)+p(BnAy)},\\
\q(BnAy)=
S_2\times
\frac{p(BnAy)}{p(ByAn)+p(BnAy)}.
}
We call $\q(\cdots)$ the QQE-renormalization of $p(\cdots)$. 
Then, as shown in the following theorem,
the QQE-renormalizations $\q(\cdots)$ are 
joint probability distributions satisfying  QQE, which are close to 
the original  joint probability distributions  $p(\cdots)$ in the sense that 
\deq{1-\frac{\q(\cdots)}{p(\cdots)}\sim \frac{|q|}{2\min(S_1,S_2)}}
if $|q|\ll 1$.  Moreover, if the original  joint probability distributions  $p(\cdots)$ 
satisfy QQE, then we have  $\q(\cdots)=p(\cdots)$.

\bTheorem\label{th:QQE-renormalization}
The following statements hold.

(i) The QQE-renormalizations $\q(AaBb), \q(BaAb)$ with $a,b=y,n$
define joint probability distributions; i.e.,  $\q(AaBb)$ and $\q(BaAb)$
satisfy
$0\le \q(AaBb), \q(BaAb)\le1 $ for all $a,b=y,n$ and
$\sum_{a,b=y,n}\q(AaBb)=\sum_{a,b=y,n} \q(BaAb)=1$.

(ii) The joint probability distributions 
$\{\q(AaBb), \q(BaAb)\mid a,b=y,n\}$ satisfy the QQE, i.e., 
\deqs{
q&=[\q(ByAy)+\q(BnAn)]-[\q(AyBy)+\q(AnBn)]\nn\\
&=[\q(AyBn)+\q(AnBy)]-[\q(ByAn)+\q(BnAy)] =0.
}

(iii) If the joint probability distributions $p(AaBb), p(BaAb)$ with $a,b=y,n$
satisfy the QQE then  $\q(AaBb)=p(AaBb)$ and $\q(BaAb)=p(BaAb)$
for all $a,b=y,n$.

(iv) The following relations hold.
\deqs{
\lefteqn{\q(AyBy)+\q(AnBn)+\q(ByAy)+\q(BnAn)}\qquad \qquad\\
&=p(AyBy)+p(AnBn)+p(ByAy)+p(BnAn),\nn\\
\lefteqn{\q(AyBn)+\q(AnBy)+\q(ByAn)+\q(BnAy)}\qquad \qquad\\
&=p(AyBn)+p(AnBy)+p(ByAn)+p(BnAy).}

(v) The following relations hold.
\deqs{
\frac{\q(AyBy)}{\q(AyBy)+\q(AnBn)}&=
\frac{p(AyBy)}{p(AyBy)+p(AnBn)},\\
\frac{\q(AnBn)}{\q(AyBy)+\q(AnBn)}&=
\frac{p(AnBn)}{p(AyBy)+p(AnBn)},\\
\frac{\q(ByAy)}{\q(ByAy)+\q(BnAn)}&=
\frac{p(ByAy)}{p(ByAy)+p(BnAn)},\\
\frac{\q(BnAn)}{\q(ByAy)+\q(BnAn)}&=
\frac{p(BnAn)}{p(ByAy)+p(BnAn)},\\
\frac{\q(AyBn)}{\q(AyBn)+\q(AnBy)}&=
\frac{p(AyBn)}{p(AyBn)+p(AnBy)},\\
\frac{\q(AnBy)}{\q(AyBn)+\q(AnBy)}&=
\frac{p(AnBy)}{p(AyBn)+p(AnBy)},\\
\frac{\q(ByAn)}{\q(ByAn)+\q(BnAy)}&=
\frac{p(ByAn)}{p(ByAn)+p(BnAy)},\\
\frac{\q(BnAy)}{\q(ByAn)+\q(BnAy)}&=
\frac{p(BnAy)}{p(ByAn)+p(BnAy)}.
}

(v) The following relations hold.
\deqs{
\frac{\q(AyBy)}{p(AyBy)}-1
=\frac{q}{2S_1-q},\\
\frac{\q(AnBn)}{p(AnBn)}-1
=\frac{q}{2S_1-q},\\
\frac{\q(ByAy)}{p(ByAy)}-1
=\frac{-q}{2S_1+q},\\
\frac{\q(BnAn)}{p(BnAn)}-1
=\frac{-q}{2S_1+q},\\
\frac{\q(AyBn)}{p(AyBn)}-1
=\frac{-q}{2S_2+q},\\
\frac{\q(AnBy)}{p(AnBy)}-1
=\frac{-q}{2S_2+q},\\
\frac{\q(ByAn)}{p(ByAn)}-1
=\frac{q}{2S_2-q},\\
\frac{\q(BnAy)}{p(BnAy)}-1
=\frac{q}{2S_2+q}.
}
\eTheorem
\bProof
The assertions can be verified straightforward calculations.
\eProof
From Theorem \ref{th:QQE-renormalization} (iv), if $S_1=S_2$ and $|q|\ll 1$, the renormalization error $(\q-p)/p$
is uniformly $|q|$, and if $S_1<S_2$ and $|q|\ll 1$, it is uniformly $|q|/(2S_1)$.

\section{Independence of personality state}

In the previous sections, we have not assumed that in the probability distribution
$\mu$ of the mind state, the personality state is independent from the 
belief state. 
According to section \ref{se:I_A} the personality state determines the type of 
how the belief state is changed by $U_A$ and $U_B$.  
It would be desirable to describe it to be independent 
from the belief state $(\al,\be)$.  

If the personality state 
depends on belief states, the same subject responds to questions in various patterns,
so that it is difficult to interpret the role of this parameter, 
and according to the number of free parameters it is likely that we can always find
a suitable input data to explain the QQE renormalized output data by our model.
On the other hand, if the personality state is independent of belief states, 
each individual subject is classified by the personality state and each
ensemble of subjects for the same experiment can be classified by the probability
distribution of the personality state.  Thus, the statistical role of the personality state
is quite clear.  

However, in this case, the number of free parameter becomes not
enough to reproduce all the QQE  renormalized output data.
Thus, the new assumption that the personality state is independent of the belief
state bring about the following interesting problems: (i) What QQE  renormalized 
output data can be reproduced by the current model?  (ii) In such output data, what are
characteristic features of the probability distribution of the personality state in 
the input data? (iii) How can we extend the role of the personality state to cover 
more QQE renormalized output data?

For the case where only personality state 
is $\ket{0}$, the model reduces to Bayesian update model, which does not have
QOE.  In our preceding paper \cite{ENTROPY}, 
we studied the model only with personality state $\ket{1}$ and showed 
that QOE and RRE hold.  In this paper, we shall further show that the model with personality state
$\ket{0},\ket{1},\ket{2}$ can account for the Clinton-Gore experiment.  

Here, we assume the independence of the personality state.
Thus, we suppose 
\begin{align}
\mu(\al,\be,\ga)&=p(\al,\be)q(\ga),\\
p(\al,\be)&=\sum_{\ga}\mu(\al,\be,\ga),\\
q(\ga)&=\sum_{\al,\be}\mu(\al,\be,\ga).
\end{align}

Then, Eqs.~\eq{JPD-1}--\eq{JPD-8} are written as follows.
\deq{
p(A{y}B{y})&=p(1,1)q(0)+p(1,1)q(2)+p(1,0)q(2).\label{eq:JPD-A1}\\
p(A{y}B{n})&=p(1,0)q(0)+p(1,0)q(1)+p(1,1)q(1).\label{eq:JPD-A2}\\
p(A{n}B{y})&=p(0,1)q(0)+p(0,1)q(1)+p(0,0)q(1).\label{eq:JPD-A3}\\
p(A{n}B{n})&=p(0,0)q(0)+p(0,0)q(2)+p(0,1)q(2).\label{eq:JPD-A4}\\
p(B{y}A{y})&=p(1,1)q(0)+p(1,1)q(2)+p(0,1)q(2).\label{eq:JPD-A5}\\
p(B{n}A{y})&=p(1,0)q(0)+p(1,0)q(1)+p(0,0)q(1).\label{eq:JPD-A6}\\
p(B{y}A{n})&=p(0,1)q(0)+p(0,1)q(1)+p(1,1)q(1).\label{eq:JPD-A7}\\
p(B{n}A{n})&=p(0,0)q(0)+p(0,0)q(2)+p(1,0)q(2).\label{eq:JPD-A8}
}

Let $p(Ay)=p(AyBy)+p(AyBn)$, etc.  Then, $p(Ay)=p(1,1)+p(1,0)$, etc.
We have
\begin{align}
p(A{y}B{y})&=p(1,1)q(0)+p(Ay)q(2).\label{eq:JPD-B1}\\
p(A{y}B{n})&=p(1,0)q(0)+p(Ay)q(1).\label{eq:JPD-B2}\\
p(A{n}B{y})&=p(0,1)q(0)+p(An)q(1).\label{eq:JPD-B3}\\
p(A{n}B{n})&=p(0,0)q(0)+p(An)q(2).\label{eq:JPD-B4}\\
p(B{y}A{y})&=p(1,1)q(0)+p(By)q(2).\label{eq:JPD-B5}\\
p(B{n}A{y})&=p(1,0)q(0)+p(Bn)q(1).\label{eq:JPD-B6}\\
p(B{y}A{n})&=p(0,1)q(0)+p(By)q(1).\label{eq:JPD-B7}\\
p(B{n}A{n})&=p(0,0)q(0)+p(Bn)q(2).\label{eq:JPD-B8}
\end{align}
Thus, if $p(Ay)\ne p(By)$ and $p(Ay)\ne p(Bn)$,
we can determine $q(0), q(1), q(2)$ by the experimental data:
\deq{
q(2)&=\frac{p(A{y}B{y})-p(B{y}A{y})}{p(Ay)-p(By)},\label{eq:JPD-D1}\\
q(1)&=\frac{p(A{y}B{n})-p(B{n}A{y})}{p(Ay)-p(Bn)},\label{eq:JPD-D2}\\
q(1)&=\frac{p(A{n}B{y})-p(B{y}A{n})}{p(An)-p(By)},\label{eq:JPD-D3}\\
q(2)&=\frac{p(A{n}B{n})-p(B{n}A{n})}{p(An)-p(Bn)}.\label{eq:JPD-D4}
}
Note that  since the output data $\{p(AaBb), p(BbAa)\mid a,b=y,n\}$
obtained from our model 
satisfy the QQ equality, the above relations are consistent,
so that $q(1)$ and $q(2)$ are uniquely determined from the experimental data 
and then we determine $q(0)$ by $q(0)+q(1)+q(2)=1$.

If we obtain the personality state $q(0),q(1),q(2)$, the belief state is determined by 
the following relations.
\begin{align}
p(1,1)&=\frac{\q(AyBy)-\q(Ay)q(2)}{q(0)},\label{eq:JPD-D5}\\
p(1,0)&=\frac{\q(AyBn)-\q(Ay)q(1)}{q(0)},\\
p(0,1)&=\frac{\q(AnBy)-\q(An)q(1)}{q(0)},\\
p(0,0)&=\frac{\q(AnBn)-\q(An)q(2)}{q(0)},\label{eq:JPD-D8}\\
p(1,1)&=\frac{\q(ByAy)-\q(By)q(2)}{q(0)},\\
p(0,1)&=\frac{\q(ByAn)-\q(By)q(1)}{q(0)},\\
p(1,0)&=\frac{\q(BnAy)-\q(Bn)q(1)}{q(0)},\\
p(0,0)&=\frac{\q(BnAn)-\q(Bn)q(2)}{q(0)}\label{eq:JPD-D12}.
\end{align}

Therefore, if $p(Ay)\ne p(By)$ and $p(Ay)\ne p(Bn)$,
and Eqs.~\eq{JPD-D1}--\eq{JPD-D12} determines the probability distributions
$\{q(0),q(1),q(2)\}$ and joint probability distributions $\{p(0,0),\ldots,p(1,1)\}$,
the QQE renormalized data $\{\q(A{y}B{y}),\ldots,\q(A{n}B{n})\}$ can be
exactly reproduced by our model.
A precise statement for the reproducibility condition is given as follows.

\bTheorem
Suppose that joint probabilities 
$p(AyBy)$, $p(AyBn)$, $p(AnBy)$, $p(AnBn)$, $p(ByAy)$,
$p(ByAn)$, $p(BnAy)$, and $p(BnAn)$ for $a,b=y,n$,
i.e., $p(Aa,Bb), p(Bb,Aa)\ge 0$, $\sum_{a,b=y,n}p(Aa,Bb)=
\sum_{a,b=y,n}p(Bb,Aa)=1$, 
are labeled as
\[
p(Ay)> p(By)\ge p(Bn)\ge p(An),
\] 
and satisfy the QQ-equality \eq{QQE}.
Then, there exist input joint probabilities  $q(0), q(1), q(2)$ and
$p(0,0), p(0,1), p(1,0), p(1,1)$ uniquely such that the model outputs
the joint probabilities $\{p(AaBb), p(Bb,Ab)\mid a,b=y,n\}$ if and
only if the following conditions hold:
\deqs{
(i)  &\quad p(AyBy)\ge p(ByAy),\\
(ii) &\quad  p(AyBn)\ge p(BnAy),\\
(iii) &\quad \frac{p(AyBn)-p(ByAn)}{p(Ay)-p(By)}
\ge\frac{p(AyBn)-p(BnAy)}{p(Ay)-p(Bn)},\\
(iv) &\quad \frac{p(ByAy)}{p(By)}\ge\frac{p(AyBy)}{p(Ay)},\\ 
(v) &\quad \frac{p(BnAy)}{p(Bn)}\ge\frac{p(AyBn)}{p(Ay)},\\ 
(vi) &\quad  \frac{p(AnBy)}{p(An)}\ge\frac{p(ByAn)}{p(By)},\\ 
(vii) &\quad \frac{p(AnBn)}{p(An)}\ge \frac{p(BnAn)}{p(Bn)}.
}

\eTheorem
\bProof
Necessity: 
Relations (i) and (ii) follow from \eq{JPD-D1} and
\eq{JPD-D2} with the assumption $p(Ay)\ge p(By)\ge p(Bn)$.
Relation (iii) follows from relations \eq{JPD-D1}, \eq{JPD-D2}
and $1-q(1)\ge q(2)$; note that the assumption $p(Ay)>p(By)$
implies the relation $p(Bn)>p(An)$. 
From \eq{JPD-B1} and \eq{JPD-B5} we have
\deqs{
\frac{p(AyBy)}{p(Ay)}&=\frac{q(0)p(1,1)}{p(Ay)}+q(2),\\
\frac{p(ByAy)}{p(By)}&=\frac{q(0)p(1,1)}{p(By)}+q(2),
}
and hence relation (iv) follows from the assumption $p(Ay)>p(By)$
and $p(1,1)\ge 0$.
Relations (v), (vi), (vii) follows similarly from conditions
$q(0)p(1,0)\ge 0, q(0)p(1,0)\ge 0$, and $q(0)p(0,0)\ge 0$,

Sufficiency: 
Suppose that $q(0), q(1), q(2), p(0,0),
\ldots, p(1,1)$ satisfy \eq{JPD-B1}, \ldots,  \eq{JPD-B8},
and $q(0)+q(1)+q(2)=1$.
It suffices to prove that they are non-negative.
We have $p(2)\ge 0$ by relation (i) and \eq{JPD-D1}.
Similarly, $p(1)\ge 0$ follows from relation (ii) and \eq{JPD-D2},
and $q(0)\ge 0$ follows from relation (iii) and $q(0)+q(1)+q(2)=1$.
Then, $p(0,0)\ge 0$ follows from relation (iv) and \eq{JPD-B1}.
The non-negativity of $p(1,0), p(0,1), p(0,0)$ follows similarly.  
\eProof

\section{Clinton-Gore poll}

In this section, we shall show that the well-known data from Clinton-Gore experiment 
can be reproduced within $\pm 0.75\%$ of errors from our model.

Consider the following data from Clinton-Gore experiment
\cite{Moo02,WB13,WSSB14}.
\begin{align}
p(AyBy)&=0.4899,\label{eq:CGE-1}\\
p(AyBn)&=0.0447,\\
p(AnBy)&=0.1767,\\
p(AnBn)&=0.2887,\\
p(ByAy)&=0.5625,\\
p(ByAn)&=0.1991,\\
p(BnAy)&=0.0255,\\
p(BnAn)&=0.2129.\label{eq:CGE-8}
\end{align}

The QQ equality is approximately satisfied
with good accuracy.
\begin{align}
q&=[p(ByAy)+p(BnAn)]-[p(AyBy)+p(AnBn)]\nn\\
&=[p(AyBn)+p(AnBy)]-[p(ByAn)+p(BnAy)] =-0.0032.
\end{align}
Thus, their QQE-renormalization $\q(Aa,Bb), \q(Bb,Aa)$ are expected to
approximate the original data  $p(Aa,Bb), p(Bb,Aa)$ with  good accuracy.

For the Clinton-Gore poll, we have
\deq{
S_1=\frac{p(AyBy)+p(AnBn)+p(ByAy)+p(BnAn)}{2}
=0.7770,\\
S_2=\frac{p(AyBn)+p(AnBy)+p(ByAn)+p(BnAy)}{2}
=0.2230,
}
and we obtain their QQE-renormalizations as follows.
\begin{align*}
\q(AyBy)&=S_1\times\frac{p(AyBy)}{p(AyBy)+p(AnBn)}
=0.4889,\\
{\rm Error}&=\frac{\q(AyBy)}{p(AyBy)}-1
=-0.0021.\ (-0.21\%),\\
\q(AnBn)&=S_1\times\frac{p(AnBn)}{p(AyBy)+p(AnBn)}
=0.2881,\\
{\rm Error}&=\frac{\q(AyBn)}{p(AyBn)}-1
=-0.0021.\ (-0.21 \%),\\
\q(ByAy)&=S_1\times\frac{p(ByAy)}{p(ByAy)+p(BnAn)}
=0.5637,\\
{\rm Error}&=\frac{\q(ByAy)}{p(ByAy)}-1
=0.0021.\ (+0.21 \%),\\
\q(BnAn)&=S_1\times\frac{p(BnAn)}{p(ByAy)+p(BnAn)}
=0.2133,\\
{\rm Error}&=\frac{\q(BnAn)}{p(BnAn)}-1
=0.0020.\ (+0.20 \%),\\
\q(AyBn)&=S_2\times\frac{p(AyBn)}{p(AyBn)+p(AnBy)}
=0.0450,\\
{\rm Error}&=\frac{\q(AyBn)}{p(AyBn)}-1
=0.0072.\ (+0.72 \%),\\
\q(AnBy)&=S_2\times\frac{p(AnBy)}{p(AyBn)+p(AnBy)}
=0.1780,\\
{\rm Error}&=\frac{\q(AnBy)}{p(AnBy)}-1
=0.0072.\ (+0.72 \%),\\
\q(ByAn)&=S_2\times\frac{p(ByAn)}{p(ByAn)+p(BnAy)}
=0.1977,\\
{\rm Error}&=\frac{\q(AnBy)}{p(AnBy)}-1
=-0.0071.\ (-0.71\%),\\
\q(BnAy)&=S_2\times\frac{p(BnAy)}{p(ByAn)+p(BnAy)}
=0.0253,\\
{\rm Error}&=\frac{\q(BnAy)}{p(BnAy)}-1
=-0.0075. (-0.75 \%).\\
\end{align*}

Under the assumption of the independence of the personality state,
from \Eq{JPD-D1} and the relation $q(0)+q(1)+q(2)=1$,
we can determine $q(0), q(1), q(2)$ by the QQE-normalized data as follows.
\begin{align}
q(2)&=0.3288,\label{eq:JPD-F1-CGR}\\\
q(1)&=0.0668,\label{eq:JPD-F2-CGR}\\
q(0)&=0.6045.\label{eq:JPD-F3-CGR}
\end{align}
We can determine $p(\al,\be)$ for $\al,\be=0,1$ by Eqs.~\eq{JPD-B1}--\eq{JPD-B8}
as follows.
\begin{align}
p(1,1)&=0.5184,\\
p(1,0)&=0.0155,\\
p(0,1)&=0.2429,\\
p(0,0)&=0.2231,
\end{align}
Note that the unity of the total probability is satisfied. 
\[
p(1,1)+p(1,0)+p(0,1)+p(0,0)=1.0000.
\]

Thus, the belief state $p(\al,\be)$ and the personality
state $q(\ga)$ are determined from the experimental data.  
Then, our quantum model with the belief state $p(\al,\be)$ and the personality
state $q(\ga)$ accurately reconstructs the QQE-renormalized data 
$\q(Aa,Bb)$, $\q(Bb,Aa)$  for $a,b=y,n$
as follows.
\begin{align*}
\q(AyBy)
&=p(1,1)q(0)+[p(1,1)+p(1,0)]q(2)
=0.4889,\\
\q(AyBn)&=p(1,0)q(0)+[p(1,1)+p(1,0)]q(1)
=0.0450,\\
\q(AnBy)&=p(0,1)q(0)+[p(0,1)+p(0,0)]q(1)
=0.1780,\\
\q(AnBn)&=p(0,0)q(0)+[p(0,1)+p(0,0)]q(2)
=0.2881,\\
\q(ByAy)&=p(1,1)q(0)+[p(0,1)+p(1,1)]q(2)
=0.5637,\\
\q(ByAn)&=p(0,1)q(0)+[p(0,1)+p(1,1)]q(1)
=0.1977,\\
\q(BnAy)&=p(1,0)q(0)+[p(0,0)+p(1,0)]q(1)
=0.0253,\\
\q(BnAn)&=p(0,0)q(0)+[p(0,0)+p(1,0)]q(2)
=0.2133.
\end{align*}
Therefore, all data of the QQR-renormalizations 
are accurately reproduced,
and we conclude that our quantum model reproduces the statistics of the 
Clinton-Gore Poll data almost faithfully  (within $\pm 0.75\%$ of error) with a {\it prior belief state $\{p(0,0),
\ldots, p(1,1)\}$ independent
of the question order}.  Thus, this model successfully corrects for the order
effect in the data to determine what in the model is the genuine distribution of the 
opinions.

\section*{Acknowledgments}
This work was partially supported by JSPS KAKENHI, Nos.~26247016 and 
17K19970.  M.O. acknowledges the support of the IRI-NU collaboration.

\medskip
\appendix

\section{Positive operator valued measures}
\label{apovm}
We restrict considerations to POVMs with a discrete domain  $X=\{x_1,...,x_N\}.
$ POVM is a map $x \to  \Pi^A(x)$, where $A$ is a symbol for the outcome variable, 
but does not denote any self-adjoint operator nor observable to be measured.
  Here, for each $x \in X,  \Pi^A(x)$ is a positive contractive
  self-adjoint operator (i.e.,  $0\le \Pi^A(x)\le I$) (called an {\it effect}), and the normalization condition 
\begin{equation}
\label{zzz}  
\sum_x \Pi^A(x)=I
\end{equation}
holds, where $I$ is the unit operator. This map defines an operator valued measure on the algebra of all subsets of $X,$
for $O \subset X, \; \Pi^A(O)= \sum_{x \in O} \Pi^A(x).$ The condition (\ref{zzz}) characterizes 
``probability operator 
valued measures''.
From (\ref{HHar}), we see that the map $x \to E^A(x)$ is a special sort of POVM, the projection valued measure - PVM.

POVMs  $\Pi^A$ represent statistics of  measurements with the outcome variable $A$
of quantum observables with the following generalization  of the Born's rule:   
$$
\Pr\{A =x\|\rho\} = \Tr[ \Pi^A(x)\rh]. 
$$ 
We remark that equality (\ref{zzz}) implies that $\sum_x \Pr\{A =x\|\rho\}=1.$

POVM does not represent state update. The latter is typically determined (non-uniquely) via representation of effects  in the form:
\begin{equation}
\label{BRULEbt}  
\Pi^A(x)=   V(x)^\star  V(x),
\end{equation}
where $V(x)$ is a linear operator in $H$; a canonical choice is $V(x)= \Pi^A(x)^{1/2}$
and every $V(x)$ is of the form  $V(x)= U(x)\Pi^A(x)^{1/2}$ with unitary operators  $U(x)$ for any $x$ by polar decomposition.
  Hence, similarly to  (\ref{HHa})
the normalization condition has  the form
$\sum_x V(x)^\star V(x) = I.$
The Born rule can be written similarly to  (\ref{eq:Born2}):   
\begin{equation}
\label{BRULEb}
\Pr\{A =x\|\rho\} = \Tr[V(x)\rh V^\star(x)]
\end{equation}
It is assumed that the post-measurement state transformation is based on the map: 
\begin{equation}
\label{TRAr}
\rho \to \cI_A(x)\rho= V(x)\rh V^\star(x), 
\end{equation}
so 
\begin{equation}
\label{TRA4}
\rho \to \rho_{\{A=x\}}= \frac{\cI_A(x)\rho}{\Tr [\cI_A(x)\rho]}.
\end{equation}

Now, we remark that the map  $x \to \cI_A(x)$ given by  (\ref{TRAr}) is a (very special) quantum instrument.
We would like to elevate the role of the use of quantum instruments, comparing with the use of just POVMs.  
An instrument provides both statistics of the measurement-outputs and the rule for the state update, but POVM should always be 
endowed with ad hoc condition (\ref{BRULEbt}). 

Finally, we remark that any instrument $\cI_A$, with the outcome variable $A$, 
generates POVM $\Pi^A$ by the rule:
\begin{equation}
\label{eq:TRA4m}
\Pi^A(x) = \cI_A(x)^\star I.
\end{equation} 
However, its state update need not have the form (\ref{TRAr}).
The general form will be give as follows.
Recall that any instrument $\cI_A$
has a family $\{M_{xj}\}_{x,j}$ of measurement operators satisfying
\deqs{\cI_A(x)\rho= \sum_{j} M_{xj}\rho M_{xj}^*.}
Then, from \Eq{TRA4m} the POVM $\Pi^A$ generated by the instrument $\cI_A$ is given by
\deq{\label{eq:POVM-I}\Pi^A(x) = \sum_{j} M_{xj}^*M_{xj}.}
Thus, Eq.~(\ref{BRULEbt}) is a special case of \Eq{POVM-I} where $V(x)=  M_{xj}$ with
$j\in\{1\}$. 

\section{Derivation of \Eq{SM-1}}
\label{se:SM}
It can be shown that for commuting observables $A$ and $B$ there exist an observable $C$
and polynomials $f$ and $g$ such that $A=f(C)$ and $B=g(C)$  \cite{VN}.  
Then, \Eq{SM-1} follows from the relations 
\deqs{\Pr\{A=x,B=y\|\rho\}
&=\Pr\{f(C)=x, g(C)=y\|\rho\}=\sum_{u:f(u)=x,g(u)=y}\Pr\{C=u\|\rho\}\\
&=\sum_{u:f(u)=x,g(u)=y}\Tr[E^{C}(u)\rho]
=\Tr[\sum_{u:f(u)=x}E^{C}(u)\sum_{v:g(v)=y}E^{C}(v)\rho]\\
&=\Tr[E^{f(C)}(x)E^{g(C)}(y)\rho]=\Tr[E^{A}(x)E^{B}(y)\rho] .}
The third last equation follows from $E^{C}(u)E^{C}(v)=0$ if $u\not=v$.

\section{Derivations of  \Eq{I_A-A} and \Eq{IA}} 
\label{ap:1}
Suppose that the object state $\ket{\psi}$ just before the measurement is arbitrarily
given, i.e., $\ket{\psi}=\sum_{\al,\be,\ga}c_{\al,\be,\ga}\ket{\al,\be,\ga}$.
Then, by the Born formula the probability distribution of the observable $A$ is defined as
\begin{align}
\Pr\{A=0| \ket{\psi}\}&=\|A^\perp\kps\|^2=\sum_{\be,\ga}|c_{0\be\ga}|^2,\\
\Pr\{A=1| \ket{\psi}\}&=\|A\kps\|^2=\sum_{\be,\ga}|c_{1\be\ga}|^2.
\end{align}
By linearity of $U_A$, it follows from Eqs. \eq{U_A-1}--\eq{U_A-4} that 
\begin{align}\label{eq:U_A}
\lefteqn{\quad U_A:\ket{\psi}\ket{00}\mapsto}\nn\\
& \sum_{\al,\be}\Big(c_{\al,\be,0}\ket{\al,\be,0}
+c_{\al,\be,1}\ket{\al,\al\Perp,1}
+c_{\al,\be,2}\ket{\al,\al,2}\Big)\ket{\al,\be}.
\end{align}

Then, 
we have
\begin{align}\label{eq:MP-X}
\lefteqn{
\left(I_{\cH}\otimes P^{M_A}(a)\right)U_A\ket{\psi}\ket{00}}\quad\nn\\
&=\sum_{\be}\Big(c_{a,\be,0}\ket{a,\be,0}+c_{a,\be,1}\ket{a,a\Perp,1}
+c_{a,\be,2}\ket{a,a,2}\Big)\ket{a,\be}
\end{align}
for $a=0,1$.
Consequently,
\begin{align*}
(I_{\cH}\otimes M_A^\perp)U_A\ket{\psi}\ket{00}
&=\sum_{\be}\Big(c_{0,\be,0}\ket{0,\be,0}+c_{0,\be,1}\ket{0,1,1}
+c_{0,\be,2}\ket{0,0,2}\Big)\ket{0,\be},
\\
(I_{\cH}\otimes M_A)U_A\ket{\psi}\ket{00}
&=\sum_{\be}\Big(c_{1,\be,0}\ket{1,\be,0}+c_{1,\be,1}\ket{1,0,1}
+c_{1,\be,2}\ket{1,1,2}\Big)\ket{1,\be}.
\end{align*}
The probabilities of obtaining the outcomes $\ba=0$ and $\ba=1$ are given by
\begin{align*}
\Pr\{\ba=0|\rho\}&=\Pr\{M_A=0|\, U_A\ket{\psi}\ket{00}\}=\|(I_{\cH}\otimes M_A^\perp) U_A \ket{\psi}\ket{00}\|^2
=\sum_{\be,\ga}|c_{0\be\ga}|^2,\\
\Pr\{\ba=1|\rho\}&=\Pr\{M_A=1|\, U_A\ket{\psi}\ket{00}\}=\|(I_{\cH}\otimes M_A)U_A\ket{\psi}\ket{00}\|^2
=\sum_{\be,\ga}|c_{1\be\ga}|^2.
\end{align*}
This shows 
\begin{align*}
\Pr\{\ba=0|\rho\}&=\Pr\{A=0|\, \ket{\psi}\},\\
\Pr\{\ba=1|\rho\}&=\Pr\{A=1|\, \ket{\psi}\}
\end{align*}
for any state $\kps$ in $\cH$.
It follows that the instrument $\cI_A$ measures the observable $A$,
i.e., 
\begin{align*}
\Pr\{\ba=0|\rho\}&=\Tr[\cI_A(0)\rh]=\Tr[A^\perp\rh],\\
\Pr\{\ba=1|\rho\}&=\Tr[\cI_A(1)\rh]=\Tr[A\rh]
\end{align*}
for any density operator $\rh$ on $\cH$.
Thus, we obtain \Eq{I_A-A}.

From Eqs.~\eq{ins-X} and  \eq{MP-X} we have
\begin{align*}
\lefteqn{
\cI_A(a)\ketbra{\psi}}\\
=&
\sum_{\be}|c_{a,\be,0}|^2\ketbra{a,\be,0}
+|c_{a,\be,1}|^2\ketbra{a,a\Perp,1}+|c_{a,\be,2}|^2\ketbra{a,a,2}\\
=&
\sum_{\be}\ket{a,\be,0}\av{a,\be,0|\psi}\av{\psi|a,\be,0}\bra{a,\be,0}
+\ket{a,a\Perp,1}\av{a,\be,1|\psi}\av{\psi|a,\be,1}\bra{a,a\Perp,1}\\
&+
 \ket{a,a,2}\av{a,\be,2|\psi}\av{\psi|a,\be,2}\bra{a,a,2}.
\end{align*} 
for $a=0,1$.
By linearity of $\cI_A(a)$ we conclude
\begin{align}
\cI_A(a)\rh&=
\sum_{\be}\ket{a,\be,0}\av{a,\be,0|\rho|a,\be,0}\bra{a,\be,0}\\
&+\ket{a,a\Perp,1}\av{a,\be,1|\rho|a,\be,1}\bra{a,\Perp,1}+
 \ket{a,a,2}\av{a,\be,2|\rho|a,\be,2}\bra{a,a,2}.\nn
\end{align} 
for any density operator $\rho$ on $\cH$ and
$a=0,1$.
Thus, we obtain \Eq{IA}.

\section{Derivation of \Eq{Affine-IA}
}\label{ap:2}
From \Eq{IA},
we have
\begin{align*}
\cI_A(0)\ket{\al,\be,\ga}\bra{\al',\be',\ga'}
= &
\sum_{\be''}\ketbra{0,\be'',0}\ket{\al,\be,\ga}\bra{\al',\be',\ga'}\ketbra{0,\be'',0}\\
&
+\Tr[(A^\perp\otimes\ketbra{1})\ket{\al,\be,\ga}\bra{\al',\be',\ga'}]\, \ketbra{010}\\
& +
\Tr[(A^\perp\otimes\ketbra{2})\ket{\al,\be,\ga}\bra{\al',\be',\ga'}]\, \ketbra{002}\\
= &
\ketbra{0,\be,0}\ket{\al,\be,\ga}\bra{\al',\be',\ga'}\ketbra{0,\be,0}\\
&
+\bra{\al',\be',\ga'}\ketbra{0}\otimes I_2 \otimes\ketbra{1})\ket{\al,\be,\ga}\, \ketbra{010}\\
&
+\bra{\al',\be',\ga'}\ketbra{0}\otimes I_2\otimes\ketbra{2})\ket{\al,\be,\ga}\, \ketbra{002}\\
= &
\de_{\al}(0)\de_{\al'}(0)\de_{\be}(\be')\de_{\ga}(0)\de_{\ga'}(0)\ketbra{0,\be,0}\\
&
+
\de_{\al}(0)\de_{\al'}(0)\de_{\be}(\be')\de_{\ga}(1)\de_{\ga'}(1) \ketbra{010}\\
&
+
\de_{\al}(0)\de_{\al'}(0)\de_{\be}(\be')\de_{\ga}(2)\de_{\ga'}(2) \ketbra{002}.
\end{align*}
It follows that if $\av{\al,\be,\ga|\al',\be',\ga'}=0$ then 
\[
\cI_A(0)\ket{\al,\be,\ga}\bra{\al',\be',\ga'}=0.
\]
Similarly,
if $\av{\al,\be,\ga|\al',\be',\ga'}=0$ then
\begin{align*}
\cI_A(a)\ket{\al,\be,\ga}\bra{\al',\be',\ga'}&=0,\\
\cI_B(b)\ket{\al,\be,\ga}\bra{\al',\be',\ga'}&=0
\end{align*}
for any $a,b$.
Therefore, for any density operator $\rh$ we have
\begin{align*}
\cI_A(a)\rh&=\sum_{\al,\be,\ga}\mu(\al,\be,\ga)\cI_A(a)\de_{(\al,\be,\ga)}=\cI_A(a)\rh',
\\
\cI_B(b)\rh&=\sum_{\al,\be,\ga}\mu(\al,\be,\ga)\cI_B(b)\de_{(\al,\be,\ga)}=\cI_B(b)\rh',
\end{align*}
where $\mu(\al,\be,\ga)=\av{\al,\be,\ga|\rh|\al,\be,\ga}$, and 
\deqs{\label{eq:Affine-IC}
\rh'=\sum_{\al,\be,\ga} \mu(\al,\be,\ga)\de_{(\al,\be,\ga)}.
}
Thus, we obtain \Eq{Affine-IA}. 

\section{On Postulate 5}
\label{apostulate5}
As we know, a density operator $\rho$ can be decomposed into probabilistic mixtures of pure states in various ways.
We show that Postulate 5 is invariant with respect to such decompositions. 
 
Suppose we conduct an experiment on 100 people. Then, we can suppose subject $j=1,\ldots,100$ is in a pure state $\kps_j$
and the ensemble is described by the mixed state  $\rho_1=(1/100)\sum_j \vert \psi_j\rangle \langle \psi_j\vert.$
We conduct another experiment on 200 people 
and $\rho_2=(1/200)\sum_k  \vert \phi_k \rangle \langle \phi_k\vert$.
The third experiment is conducted on 300 people.
and $\rho_3=(1/300)\sum_l \vert \xi_l \rangle \langle \xi_l  \vert$.
Then,
$$
P(AaBb\|\rho_1)=(1/100)\sum_j P(AaBb\|\kps_j),
P(BbAa\|\rho_1)=(1/100)\sum_j P(BbAa\|\kps_j),
$$
$$
P(AaBb\|\rho_2)=(1/200)\sum_j P(AaBb\|\kph_j),
P(BbAa\|\rho_2)=(1/200)\sum_j P(BbAa\|\kph_j),
$$
$$
P(AaBb\|\rho_3)=(1/300)\sum_j P(AaBb\|\kxi_j),
P(BbAa\|\rho_3)=(1/300)\sum_j P(BbAa\|\kxi_j).
$$
It follows that if $p\rho_1+p^\prime\rho_2=\rho_3,$
where $p,  p^\prime >0$,  $p^\prime=1-p.$
Then
$p P(AaBb\|\rho_1) + p^\prime P(AaBb\|\rho_2)=P(AaBb\|\rho_3).$
This is consistent with
$P(AaBb\|\rho)=\Tr[\cI_B(b)\cI_A(a)\rho].$
This also holds for any longer sequences and the joint
probability distribution  depends only on $\rho$, but it is independent of
its decomposition into pure states, orthogonal or not.

\end{document}